\documentclass[prl,twocolumn,english,superscriptaddress,longbibliography]{revtex4-2}

\usepackage{graphicx}
\usepackage{float}
\usepackage[dvipsnames]{xcolor}
\usepackage{bm,amsmath,amssymb}
\usepackage{soul}
\usepackage[T1]{fontenc}

\usepackage[unicode=true,pdfusetitle,
 bookmarks=true,bookmarksnumbered=false,bookmarksopen=false,
 breaklinks=false,pdfborder={0 0 0},pdfborderstyle={},backref=false,colorlinks=true]
 {hyperref}
\usepackage{braket}
\usepackage{xfrac}
\usepackage{verbatim}
\usepackage{lipsum}
\usepackage{cancel}
\usepackage{relsize}
\usepackage{placeins} 
\usepackage{amstext}
\usepackage{bbold}
\usepackage{esint}
\usepackage{tikz}
\usepackage{soul}

\newcommand{\nbar}[0]{\bar{n}}

\newcommand{\rs}[1]{\textcolor{black}{#1}}

\newcommand{\kz}[1]{\textcolor{black}{#1}}

\newcommand{\refeo}[1]{\textcolor{black}{#1}}
\newcommand{\refet}[1]{\textcolor{black}{#1}}

\begin{document}

\title{Dissipative preparation and stabilization of many-body quantum states in a superconducting qutrit array}

\author{Yunzhao Wang}
\affiliation{Department of Physics, Washington University, St.\ Louis, Missouri 63130}
\author{Kyrylo Snizhko}
\affiliation{Institute for Quantum Materials and Technologies, Karlsruhe Institute of Technology, 76021 Karlsruhe, Germany}
\affiliation{Univ. Grenoble Alpes, CEA, Grenoble INP, IRIG, PHELIQS, 38000 Grenoble, France}
\author{Alessandro Romito}
\affiliation{Department of Physics, Lancaster University, Lancaster LA1 4YB, United Kingdom}
\author{Yuval Gefen}
\affiliation{Department of Condensed Matter Physics, Weizmann Institute of Science, Rehovot, 76100 Israel}
\author{Kater Murch}
\affiliation{Department of Physics, Washington University, St.\ Louis, Missouri 63130}

\date{\today}

\begin{abstract}
We present and analyze a protocol for driven-dissipatively preparing and stabilizing \refeo{a manifold of quantum manybody entangled states} with symmetry-protected topological order.
 Specifically, we consider the experimental platform consisting of superconducting transmon circuits and linear microwave resonators. We perform theoretical modeling of this platform via pulse-level simulations based on physical features of real devices. In our protocol, transmon qutrits are mapped onto spin-1 systems. The qutrits' sharing of nearest-neighbor dispersive coupling to a dissipative microwave resonator enables elimination of state population in the $S^\mathrm{total}=2$ subspace for each adjacent pair, and thus, the stabilization of the manybody system into the Affleck, Kennedy, Lieb, and Tasaki (AKLT) state \refeo{up to the edge mode configuration}. We also analyze the performance of our protocol as the system size scales up to four qutrits, in terms of its fidelity as well as the stabilization time. Our work shows the capacity of driven-dissipative superconducting cQED systems to host robust and self-corrected quantum manybody states that are topologically non-trivial.
\end{abstract}

\maketitle

\section{Introduction}


Dissipation is usually viewed as an undesirable process in handling quantum information, it destroys quantum coherence and should therefore be removed by quantum error correction. However, dissipative processes can also contribute novel elements for quantum information processing when controlled and engineered \cite{Verstraete2009,Plenio1999}. 
One such application is in preparing a quantum manybody system into the ground state of a generic manybody Hamiltonian \cite{Harrington2022}. 
This kind of preparation is typically achieved in one of \refeo{three} ways. One can engineer the system with the interactions of a certain Hamiltonian and relax the system towards the ground state \cite{Mazurenko2017}. However, the appropriate interactions or relaxation may not be generally achievable, or sufficiently low temperatures may pose a challenge. Alternatively, one can prepare a manybody entangled state using adiabacity \cite{Farhi2001,Kadowaki1998}, where a system is initialized in a trivial ground state and the Hamiltonian is slowly tuned to adiabatically produce the manybody ground state. Here, high fidelity requires slow evolution and an absence of excess dissipation that induce quantum jumps between states. \kz{Finally, one can start with a trivial state and implement a time-dependent Hamiltonian that will rotate the state into the desired target state, not insisting on following the instantaneous ground state, e.g. by sequential unitary operations \refeo{\cite{Schoen2005}}. However, determining and implementing the appropriate Hamiltonian with robustness to other sources of dissipation can be a challenging, if not impossible task.  The limitations of these three} approaches motivate the investigation of driven dissipative methods in preparing and stabilizing a manybody system in a non-trivial state. Here, by designing the dissipative terms in the system Lindbladian \cite{Lindblad1976}, the desired manybody state can be reached and stabilized as the fixed point of the resulting dynamics \cite{Kraus2008}.  \kz{Due to the intimate connection between disspation and measurement \cite{Roy2020}, this approach can be viewed as a type of \emph{blind steering through measurement} or equivalently \emph{autonomous feedback}.}

First proposed by Affleck, Kennedy, Lieb, and Tasaki (AKLT) in 1987, the AKLT state (Fig.~\ref{fig1}(a)) is a prototypical example of the Haldane phase \cite{Affleck1987} with a symmetry-protected topological order. It works as a resource state for measurement-based quantum computation \cite{Gross2007,Brennen2008}, 
and can be efficiently represented by matrix-product states (MPS) \cite{Perez2007}. 
Since the MPS representation efficiently describes a large variety of low-energy states of manybody Hamiltonians, protocols that can produce the AKLT state may be generalized for a range of applications.  Compared with the typical preparation method of the AKLT state based on its matrix product representation via postselection \cite{Kaltenbaek2010,Chen2022}, or based on sequential unitary gates \cite{Schoen2005} and assisted by measurements \cite{Smith2022}, driven-dissipative methods create the manybody state with robustness and self-correcting features. Here, the system coherence can last much longer than the lifetime of a single component. Prior proposals have addressed possible implementation in ion trap and cold atom systems \cite{Sharma2021,Zhou2021}.
\refeo{Here, we focus on the ALKT states under open boundary conditions, where there is a four-fold degenerate ground state. Our protocol stabilizes this subspace of states in the superconducting transmon platform.} 

\refeo{The superconducting circuit QED (cQED) platform, which we consider, provides a versatile methodology for controlling superconducting artificial atoms and their interaction with electromagnetic cavities 
\cite{Blais2021}. In recent years, the platform has been a mainstream track for developing quantum processors \cite{Kjaergaard2020}. Superconducting NISQ processors are at the forefront of the quest for quantum advantage \cite{Arute2019} and quantum error correction \cite{Acharya2023}. Beyond coherent manipulation, cQED allows for designing driven-dissipative dynamics, which has been shown to enable stabilization of single-body states \cite{Murch2012,Lu2017,Grimm2020,Leghtas2015,Valenzuela2006,Magnard2018,Holland2015,Geerlings2013}, two-body states \cite{Shankar2013, Kimchi2016, Brown2022,Ma2019_sl,Ma2021}, as well as many-body entangled states \cite{Ma2019}.}



In this article, we propose a scheme to dissipatively stabilize the 1-dimensional AKLT state consisting of spin-1 particles on such a platform. 
As is presented in Fig.~\ref{fig1}, a spin-1 chain can be realized with an array of superconducting transmon circuits \cite{Koch2007}, where each spin-1 particle is identified with the lowest three energy levels as a qutrit \cite{Blok2021}. 
Here the dissipative element is provided by autonomous feedback \cite{Shankar2013} from reservoir engineering.  
With two qutrits both coupled to a microwave resonator, local drives combined with cavity dissipation pump the qutrit pair into the subspace where their total spin $S_\mathrm{total}\in \{1,0\}$. 
In stabilization of the ground state of a frustration-free Hamiltonian, the manybody entangled state is achieved by applying such two-body dissipation terms simultaneously on each nearest neighbor pair as the system size scales up \cite{Roy2020}.
We thus demonstrate the viability of preparing and stabilizing a \refeo{4-dimensional subspace of} weakly entangled manybody states, the AKLT states, within devices of superconducting transmon qutrits, linear microwave resonators, and specific microwave drives. 

This work is structured as follows. In Section II, we introduce the physical models to describe the experimental platform and review the entanglement stabilizing \kz{scheme via autonomous feedback for the case of two qubits}. In Section III, we propose a driven-dissipative approach to stabilize the AKLT state in a superconducting qutrit array and analyze its scalability in terms of state preparation time and steady-state fidelity. Section IV discusses future extensions and concludes our article.

\section{physical models}

\subsection{Strong dispersive regime} \label{sec2a}

\begin{figure}[h]
\centering
\includegraphics[width = 0.5\textwidth]{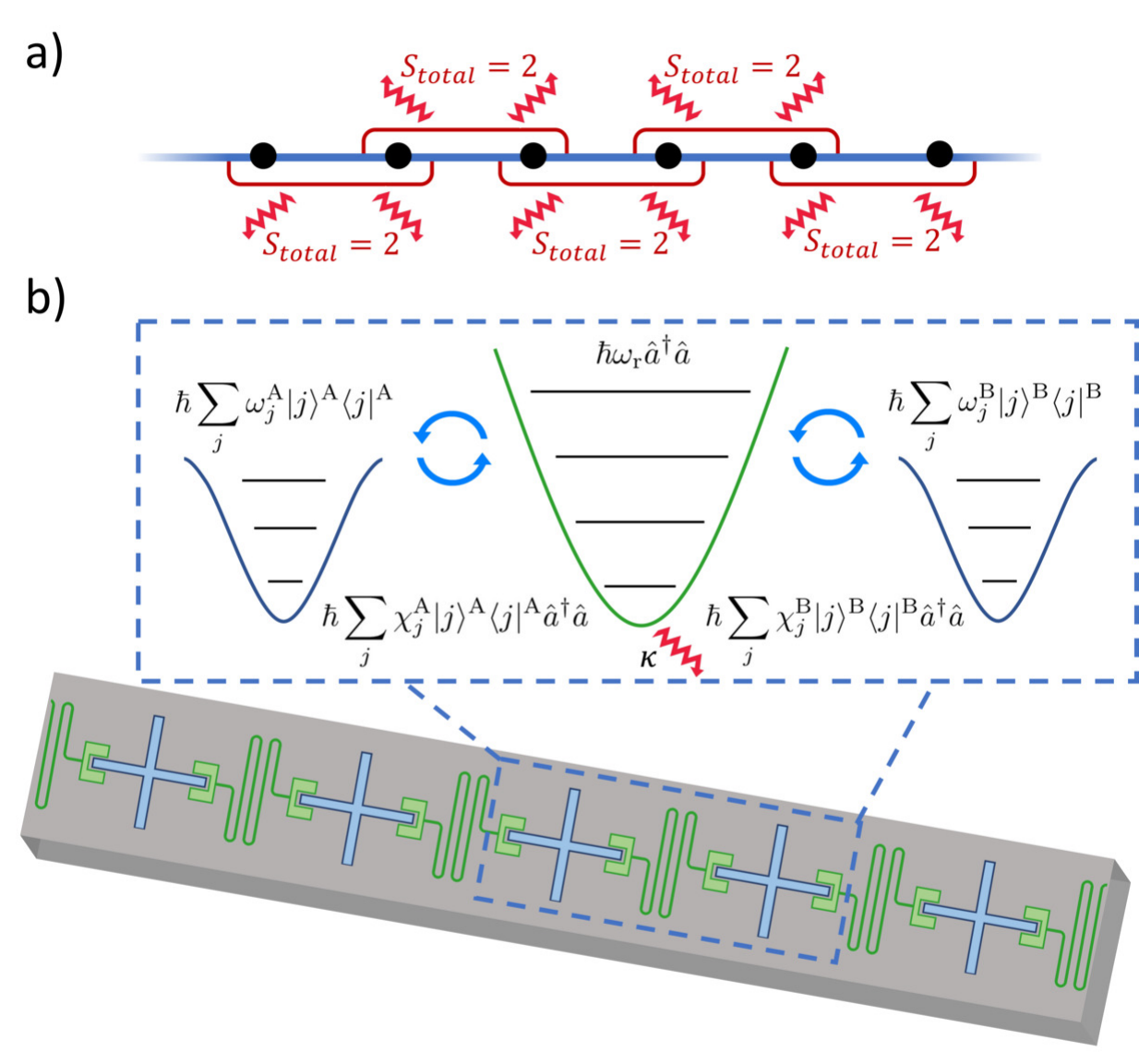}

\caption{\label{fig1}  {\bf \kz{Schematic diagram of the AKLT chain and proposed device}}. 
(a) The AKLT chain as represented in the form of a spin-1 chain, where neighboring pairs of spin-1 particles are \kz{dissipativley} excluded from the $S_\mathrm{total}=2$ manifold \kz{(indicated as red squiggly arrows)}. 
(b) Sketch of the proposed one-dimensional superconducting transmon array, with shared resonators between nearest neighbors. Using this platform we present a protocol for the dissipative preparation and stabilization of the system in the AKLT subspace.}
\end{figure}

A spin-1 one-dimensional chain can be realized experimentally by identifying each spin-1 with the lowest energy levels of a superconducting transmon circuit \cite{Bianchetti2010,Koch2007,Goss2022,Blok2021}.
\refeo{Such encoding between the spin-1 states and the native transmon states can be achieved in a simple and direct way. The eigenstates for the $z$ component of the spin can be written as $|S=1,S_z=\pm 1,0\rangle$, and the energy levels of the superconducting transmon qutrit can be written as $|g\rangle$, $|e\rangle$ and $|f\rangle$. Here we encode $|S=1,S_z=-1\rangle$ to $|g\rangle$, $|S=1,S_z=0\rangle$ to $|e\rangle$ and $|S=1,S_z=1\rangle$ to $|f\rangle$.} 
As is shown in Fig.~\ref{fig1}(b), each two adjacent transmons aligned in a 1D array are coupled to a common microwave resonator, \kz{which is coupled quasi-locally to a dissipative environment.}
A transmon qutrit coupled with a linear cavity can be described by the generalized Jaynes-Cummings Hamiltonian, 
\begin{equation}
\begin{split}
\mathrm{\hat{H}_{JC}}  
&
=\hbar \omega_\mathrm{r} ^0 \hat{a}^\dag \hat{a}+\hbar \sum_{j}  \omega_j ^0 | j \rangle\langle j|
\\
&+(\hbar \sum_{j} g_{j,j+1}| j \rangle\langle j+1|\hat{a}^\dag + h.c. ),
\end{split}
\end{equation}
with rotating wave approximation applied and qutrit parameters approaching the transmon limit \cite{Koch2007}. Here \rs{$\hat{a}^\dagger ( \hat{a}$)} is the cavity photon creation (annihilation) operator, $\ket{j} (\hbar \omega_j^0)$ are the energy eigenstates (eigenenergies) of the transmon, $\omega_\mathrm{r} ^0 /2\pi$ is the cavity resonance frequency and $g_{j,j+1}$ are the cavity-transmon couplings.
In the dispersive limit \cite{Blais2004,Wallraff2004,Blais2020}, the detunings between the cavity frequency and the qutrit transition frequencies are large compared to the coupling strength such that $|\omega_j-\omega_\mathrm{r}|/g_{j,j+1} \gg 1$. In this case, the system Hamiltonian can be approximated by
\begin{equation}
\mathrm{\hat{H}}=\hbar\omega_\mathrm{r} \hat{a}^\dag \hat{a}+\hbar \sum_{j}  \omega_j | j \rangle\langle j|+ \hbar \sum_{j} \chi_j |j\rangle \langle j|\hat{a}^\dag \hat{a},
\end{equation}
where $\omega_\mathrm{r}/2\pi$ and $ \omega_j/2\pi $ are the new cavity and transmon frequencies which have been renormalized due to the coupling.
The dispersive interaction energies, $\hbar \chi_j$, 
 shift the cavity resonance frequencies by $\chi_j/2\pi$ depending on the state of the qutrit $ |j\rangle $.
Similarly, for a cavity populated with $n$ photons, the transmon transition frequencies $(\omega_{j+1}-\omega_{j})/2\pi$ are also shifted by $(\chi_{j+1}-\chi_j)n/2\pi$, which is proportional to the photon number $n$.

\kz{As an example of the strong-dispersive regime in the context of qubits,} Fig.~\ref{fig2I}(a) displays the cavity spectrum when the dispersive interaction with a qubit shifts its resonance frequency. The strong dispersive regime \cite{Gambetta2006,Schuster2007}  occurs when the cavity linewidth $\kappa$ is much smaller than the dispersive shifts $\chi_j$, leading to well-resolved cavity spectral peaks.
In turn, the qubit spectrum is split depending on the cavity photon number \cite{Gambetta2006}, and thus the statistics of the field can be resolved \cite{Schuster2007}. \kz{The strong dispersive regime has proved to be useful in a range of different experimental tasks. For example, because of the large shifts of the cavity spectrum, quantum non-demolition measurement can be performed ``selectively'' between one state and its orthogonal subspace \cite{Wang2022}. In the context of multiple qubits, this regime has been used to demonstrate a coherent entangling gate between non-interacting qubits \cite{Blumenthal2021,Lewalle2022}.} 
\refet{In addition, the strong dispersive regime enables control and entanglement of quantum states encoded in cavity modes \cite{Su2022_1,Su2022_2,Su2022_3}.}
In this proposal, \kz{we will harness the strong dispersive regime to create autonomous feedback on the quantum state of two nearest-neighbor qutrits, the basic idea of which is introduced below with a comparatively simpler model of nearest-neighbor qubits. }

\subsection{Autonomous feedback scheme}
\label{section2B}

The purpose of our proposal is to harness dissipation engineering to prepare and stabilize a particular subspace on a chain of spin-1 (qutrits).  The underlying mechanism is similar to the one used in a  recent work focusing on the stabilization of entangled states of qubits \cite{Shankar2013}. Hence, in this Section, we review the protocol in the simpler 2-qubit scenario, before extension to the spin-1 system as presented in Section \ref{section3B}.

\begin{figure}[h]
\centering
\includegraphics[width = 0.5\textwidth]{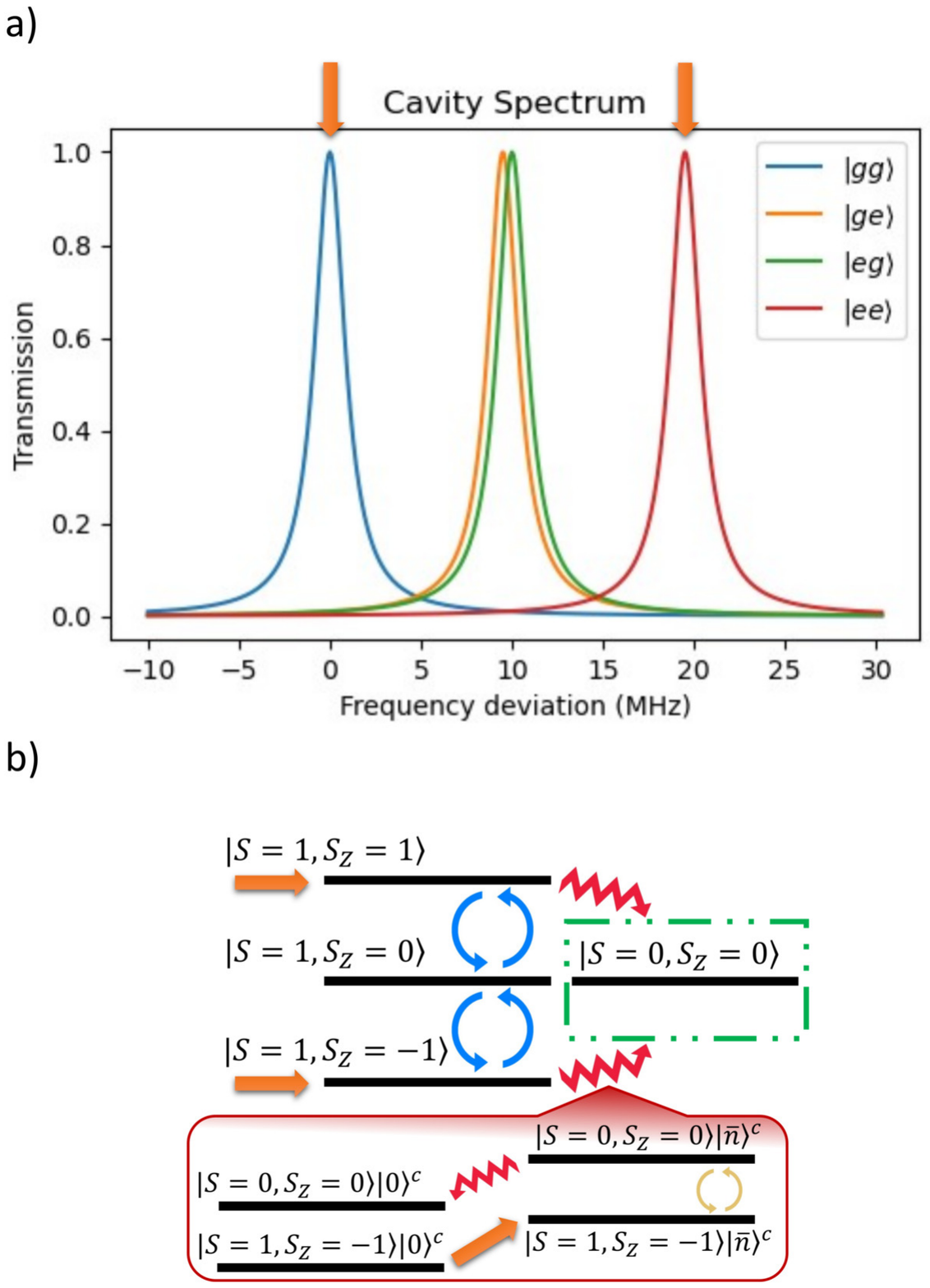}
\caption{\label{fig2I}  {\bf Protocol to prepare a two-qubit Bell state}. As introduced in \cite{Shankar2013}, a dissipative protocol can prepare and stabilize a two-qubit Bell state. (a) The cavity spectrum shifted by different two-qubit states, with two cavity probes applied at the frequencies marked by the orange arrows. (b) Hilbert space engineering for the Bell state stabilization scheme \cite{Shankar2013}. The orange arrows mark the states measured by the two cavity probes shown in (a). \kz{The blue arrows represent the ``0-photon drive'', which comes into resonance with the qubits when there are 0 photons in the cavity.} The green rectangle encircles the stabilized state, and the zigzag red arrows are dissipative processes steering into the stabilized state; the mechanism of which is shown in the inset circled by the red rectangle. This shows the realization of the jump operator from state $|gg\rangle$ and state $|ee\rangle$ to the stabilized state. The red arrows indicate populating and decaying of the cavity photons, and the yellow arrows denote the ``$n$-photon drive''.}
\end{figure}

The primary idea of stabilizing an entangled state by autonomous feedback was put forward by Leghtas et. al. for a two-qubit Bell state \cite{Leghtas2013}, and was then experimentally realized in the system of two superconducting qubits \cite{Shankar2013}, as well as with trapped ions \cite{Cole2022,Horn2018,Lin2013}. Representing a spin-half particle as a two-level system denoted by $|g\rangle$ and $|e\rangle$, Fig.~\ref{fig2I}(a) depicts the qubit-state-dependent cavity spectrum and  Fig.~\ref{fig2I}(b) demonstrates Hilbert space engineering in this dissipative stabilization scheme. For a linear cavity dispersively coupled with both qubits A and B, the system Hamiltonian with rotating wave approximation 
\begin{equation}
\begin{split}
\mathrm{\hat{H}}&=\hbar\omega_\mathrm{r} \hat{a}^\dag \hat{a}+ \hbar \omega_\mathrm{A}  \frac{\hat{\sigma}_z^\mathrm{A}}{2} +\hbar  \omega_\mathrm{B} \frac{\hat{\sigma}_z^\mathrm{B}}{2}
\\
&+\hbar g_\mathrm{A}(\hat{\sigma}_+^\mathrm{A} \hat{a} + \hat{\sigma}_-^\mathrm{A} \hat{a}^\dagger )+\hbar g_\mathrm{B}(\hat{\sigma}_+^\mathrm{B} a + \hat{\sigma}_-^\mathrm{B} \hat{a}^\dagger )
\end{split}
\end{equation}
becomes
\begin{equation}
\label{eqn:disp_qubit}
\mathrm{\hat{H}_{eff}}=\hbar \chi_\mathrm{A} \frac{\hat{\sigma}_z^\mathrm{A}}{2}  \hat{a}^\dag \hat{a}+\hbar \chi_\mathrm{B}  \frac{\hat{\sigma}_z^\mathrm{B}}{2}  \hat{a}^\dag \hat{a},
\end{equation}
in the rotating frame for qubits and the cavity, and applying the dispersive limit \refeo{(see Appendix~\ref{app:a})}.
Here $\hbar \omega_\mathrm{A(B)}$ is the transition energy of qubit A(B), $g_\mathrm{A(B)}$ is the coupling strength between the cavity and qubit A(B), and $\hbar \chi_\mathrm{A(B)}$ is the interaction energy between the cavity and qubit A(B). This indicates a shift in the cavity resonance frequency equivalent to the addition of the shifts from qubits A and B, shown in Fig.~\ref{fig2I}(a). 

The cavity is driven at $\omega_\mathrm{r}-(\chi_\mathrm{A}+\chi_\mathrm{B})/2$ and $\omega_\mathrm{r}+(\chi_\mathrm{A}+\chi_\mathrm{B})/2$, corresponding to the resonance spectrum peaks for two-qubit states $| g g \rangle $ and $| e  e \rangle $. 
Thus, whenever the qubits are in \kz{either} $| g g \rangle $ or $| e e \rangle $, the cavity photon population ramps up to an average number of $\bar{n}$. Otherwise, the cavity photon number exponentially decays to zero assuming that $\chi\gg\kappa$ limit \refeo{(See Appendix~\ref{app:b})}. 
When the cavity is populated with $n$ photons, the transition energies for qubits A and B are shifted as $\hbar \omega_\mathrm{A}^\prime=\hbar \omega_ \mathrm{A}+\hbar \chi_\mathrm{A}n$ and $\hbar \omega_\mathrm{B}^\prime=\hbar \omega_ \mathrm{B}+\hbar \chi_\mathrm{B} n$. 

Hence, we can consider two types of single-qubit Rabi drives on both qubits. The ``0-photon drive'' is applied at $\omega_\mathrm{A(B)}$ with Rabi frequency $\Omega^{(0)}$ while the ``$n$-photon drive'' is applied at $\omega_\mathrm{A(B)}+\chi_\mathrm{A(B)}n $ with Rabi frequency $\Omega^{(n)}$, given that $n \approx \bar{n}$. 
Therefore, the former qubit drive is in resonance with the cavity unpopulated while the latter requires a component into the photon number eigenstate of $n$. \refeo{The effective Hamiltonian for those continuous drives are given by}

\begin{equation}
\begin{split}
&\hat{H}_\mathrm{eff}^{(0)} \propto\hat{\sigma}_x^\mathrm{A} \otimes I^\mathrm{B}+I^\mathrm{A} \otimes \hat{\sigma}_x^\mathrm{B},
\\
&\hat{H}_\mathrm{eff}^{(n)}\propto\hat{\sigma}_x^\mathrm{A} \otimes I^\mathrm{B}-I^\mathrm{A} \otimes \hat{\sigma}_x^\mathrm{B},
\end{split}
\end{equation}

applied at their corresponding resonance frequencies.
We will now denote the cavity state with average photon number $\nbar$ in the rotating frame of its driving frequency can be denoted as $|\nbar\rangle ^\mathrm{C}$. \refet{We also denote that $| \phi_- \rangle = (|eg\rangle - |ge\rangle)/\sqrt{2}$ and $| \phi_+ \rangle = (|eg\rangle + |ge\rangle)/\sqrt{2}$.} When the qubit--cavity system is in $| g g \rangle | 0 \rangle ^\mathrm{C}$ or $| e  e \rangle| 0 \rangle ^\mathrm{C}$, the cavity starts to populate with photons and is driven to a coherent state $| g  g \rangle | \bar{n} \rangle ^\mathrm{C}$ or $| e  e \rangle | \bar{n} \rangle ^\mathrm{C}$.
At this moment, the ``$n$-photon drives'' come into resonance, rotating state $| g  g \rangle| \bar{n} \rangle ^\mathrm{C}$ and $| e e \rangle | \bar{n} \rangle ^\mathrm{C}$ into state $| \phi_- \rangle | \bar{n} \rangle ^\mathrm{C}$. 
Once leaving the two-qubit subspace spanned by states $| gg \rangle $ and $| ee \rangle $, with the system state in $| \phi_- \rangle | \bar{n}\rangle ^\mathrm{C}$, the cavity probes are no longer in resonance. Subsequently, the cavity photon population decays into $| \phi_- \rangle | 0 \rangle ^\mathrm{C}$, setting the ``$n$-photon drive'' off-resonant again. 
When $\Omega^{(n)}$ has the same scale as the cavity linewidth $\kappa $, the ``$n$-photon drive'' combined with cavity probes drives the two-qubit state unidirectionally from state $| g  g \rangle | 0 \rangle ^\mathrm{C} $ and $| e  e \rangle | 0 \rangle ^\mathrm{C}$ to the target Bell state $| \phi_- \rangle | 0 \rangle ^\mathrm{C}$. 
As is shown in the inset of Fig.~\ref{fig2I}(b), the effect of such an autonomous feedback process is similar to that of quantum jump operators $| \phi_- \rangle \langle gg|$ and $| \phi_- \rangle \langle ee|$, which occur at a rate proportional to $\kappa$. 

Aside from the above feedback loop, the ``0-photon drive'' is applied to induce rotations between state $| \phi_+ \rangle | 0 \rangle^\mathrm{C}$ and state $| gg \rangle  | 0 \rangle ^\mathrm{C}$ or $| ee \rangle  | 0 \rangle ^\mathrm{C}$. Choosing $\Omega^{(0)}$ to be comparable with $\kappa$, any state of the Hilbert space is driven into the target Bell state $\{|\phi_+\rangle, | gg \rangle , | ee \rangle \}\to |\phi_-\rangle$.
\refeo{A intuitive picture of the entire stabilizing process is shown in Fig.~\ref{fig2I}(b), where we map the two energy levels of a qubit into a spin-1/2 particle. With state $|g\rangle$ encoded into $|S=1/2,S_z=-1/2\rangle$ and state $|e\rangle$ encoded into $|S=1/2,S_z=1/2\rangle$, we have $|gg\rangle$ encoded into the added spin $|S=1,S_z=-1\rangle$, $|ff\rangle$ encoded into $|S=1,S_z=1\rangle$, $|\phi_+\rangle$ encoded into $|S=1,S_z=0\rangle$ and $|\phi_-\rangle$ encoded into $|S=0,S_z=0\rangle$.}

As is shown in Fig.~\ref{fig2I}(a), when $\chi\gg\kappa$, the density of states corresponding to $|ge\rangle$ and $|eg\rangle$ is highly suppressed at the applied cavity drive frequencies.  However, if $\chi/\kappa$ is finite, as is expected in any reasonable experimental realization, a difference in dispersive shifts ($\chi_\mathrm{A} \neq \chi_\mathrm{B}$) results in different amplitudes in the tails of the Lorentzian cavity spectrum lineshapes. This difference distinguishes the $|ge\rangle$ and $|eg\rangle$ states, corresponding to a measurement of the qubits in those bases.  
This residual measurement, therefore, dephases the $|\phi_-\rangle$ state, mixing the state populations of the stabilized state $|\phi_-\rangle$ and the eliminated state $|\phi_+\rangle$, and thus reducing the fidelity of the autonomous feedback scheme. 
Considering the scaling between the rate of this residual measurement and the stabilizing rate to the target Bell state, such a reduction in fidelity can be significant for a large discrepancy between $\chi_\mathrm{A}$ and $\chi_\mathrm{B}$.
Hence, for optimal operation, the scheme requires $\chi_\mathrm{A} \simeq \chi_\mathrm{B}$.

To summarize, the qubit protocol involves a ``pump'' that drives unwanted states $\{|\phi_+\rangle, | gg \rangle , | ee \rangle \} |0\rangle^\mathrm{C}$ to $\{| gg \rangle , | ee \rangle \} |\nbar\rangle^\mathrm{C}$. This then activates a ``reset'' which drives the qubits to $|\phi_-\rangle$, and the cavity decays to $|0\rangle^\mathrm{C}$. In the language of Roy et al. \cite{Roy2020}, the protocol belongs to the class of ``shaking and steering''.

\section{proposal}

\subsection{The AKLT state}

The AKLT Hamiltonian can be obtained as
\begin{equation}
\hat{\mathrm{H}}_\mathrm{AKLT}=\sum_i[\vec{S}_i \cdot \vec{S}_{i+1}+\frac{1}{3}(\vec{S}_i \cdot \vec{S}_{i+1})^2].
\end{equation}
Here $\vec{S}_i$ is the angular momentum vector operator for the spin-1 particle on the $i_\mathrm{th}$ site.
This model was first proposed by Affleck, Kennedy, Lieb, and Tasaki (AKLT) in 1987 as an exactly solvable model exemplifying a gapped excitation spectrum \cite{Haldane1983,Affleck1989,Wei2022} and a symmetry-protected topological (SPT) order for odd-integer spin chain \cite{Wen2017, Chen2011Cla}.
The topological order of the 1-D AKLT chain is protected by the $\mathbb{Z}_2 \times \mathbb{Z}_2$ symmetry group, and can be detected by \kz{a} string order parameter \cite{Nijs1989,Kairys2022}  or characterized by the entanglement spectrum \cite{Pollmann2010}. 
The AKLT ground state is short-range entangled, can be efficiently represented via MPS, and \refeo{cannot be modified into a non-entangled product state  without closing the gap of the Hamiltonian or breaking the} $\mathbb{Z}_2 \times \mathbb{Z}_2$ symmetry \cite{Wei2022}. 
Also, the computation capability of \kz{an open} AKLT chain as a quantum wire in measurement-based quantum computation is shared by all states in the $\mathbb{Z}_2 \times \mathbb{Z}_2$ symmetry-protected topological phase \cite{Stephen2017}.

\refet{Under periodic boundary conditions (PBC), $\hat{\mathrm{H}}_\mathrm{AKLT}$ has a unique ground state---the AKLT state.}
\refet{With open boundary conditions (OBC), though, the ground state becomes 4-fold degenerate, with two fractionalized degrees of freedom emerging on each of the two boundaries.}
\refet{The AKLT states with OBC can be explicitly written in the matrix product form \cite{Schollwock2011,Chen2022}}
\begin{equation}
|\mathrm{AKLT}\rangle = \sum_{\{s\}}
\psi(s_1,s_2,...,s_N)|s_1 s_2 ... s_N\rangle,
\end{equation}
\refet{where $s_i\in\{+,-,0\}$ are the three single particle states $|S=1,S_z=1\rangle$, $|S=1,S_z=-1\rangle$, and $|S=1,S_z=0\rangle$ for the spin-1 particles in the array, and the wave function is given by
\begin{equation}
\psi(s_1,s_2,...,s_N) = [{b_A^l}^T A^{s_1}A^{s_2}...A^{s_N}b_A^r].
\end{equation}
Here the matrices $A^+$, $A^0$ and $A^-$ are represented by}
\begin{equation}
\begin{split}
    A^+ =&\begin{pmatrix}0&\sqrt{\frac{2}{3}}\\0&0\end{pmatrix},     
    A^- =\begin{pmatrix}0&0\\-\sqrt{\frac{2}{3}}&0\end{pmatrix}, 
        \\
    &A^0 =\begin{pmatrix}-\frac{1}{\sqrt{3}}&0\\0&\frac{1}{\sqrt{3}}\end{pmatrix},
\end{split}
\end{equation}
\refet{and the boundary vectors $b_A^l$ and $b_A^r$ represent the edge spin-1/2 modes and choose a specific state out of the 4-dimensional AKLT manifold. }
\refet{In our work, we achieve stabilization into this 4-dimensional manifold with OBC, creating the AKLT state up to a boundary configuration.}
\refet{The non-trivial SPT order of the AKLT chain can be revealed by those edge modes.}
\refet{The edge states} are protected by symmetry, which means that their degeneracy can resist local perturbations that do not break the corresponding symmetry \cite{Pollmann2012}. 
For a better understanding of the edge states, one can visualize the AKLT state on a spin-1/2 chain.  In this case, the AKLT state can be obtained by \kz{dividing the chain into adjacent spin-singlet pairs, and then projecting the Hilbert space of each pair} into the spin-triplet subspace. This is the approach that is followed in optical systems for the preparation of the AKLT state for measurement-based quantum computation \cite{Kaltenbaek2010}. Such a representation views the edge modes as unpaired spin-1/2 particles.


To represent the AKLT state in a more relevant way to the driven-dissipative protocol, its parent Hamiltonian can also be written in the form of quasi-local projectors, 
\begin{equation}
\label{eqn:AKLTproj}
\hat{\mathrm{H}}_\mathrm{AKLT}=\sum_{i} \hat{P}^{S=2}_{i,i+1}.
\end{equation}
Here $\hat{P}^{S=2}_{i,i+1}$ projects a pair of neighboring spin-1 particles (the $i_\mathrm{th}$ and the $(i+1)_\mathrm{th}$) onto the subspace where their total spin equals 2, thereby adding an energetic cost to the $S_\mathrm{total}=2$ subspace. Since the AKLT Hamiltonian is frustration-free \cite{Wei2022}, its ground state can be reached by \kz{independently driving each adjacent pair of qutrits} into the $S_\mathrm{total}\in \{0,1\}$ subspace. The AKLT state can then be obtained whenever the projection onto $S_\mathrm{total}=2$ is eliminated for each adjacent pair of sites, which is \kz{schematically indicated} in Figure \ref{fig1}(a).

\begin{figure}[h]
\centering
\includegraphics[width = 0.5\textwidth]{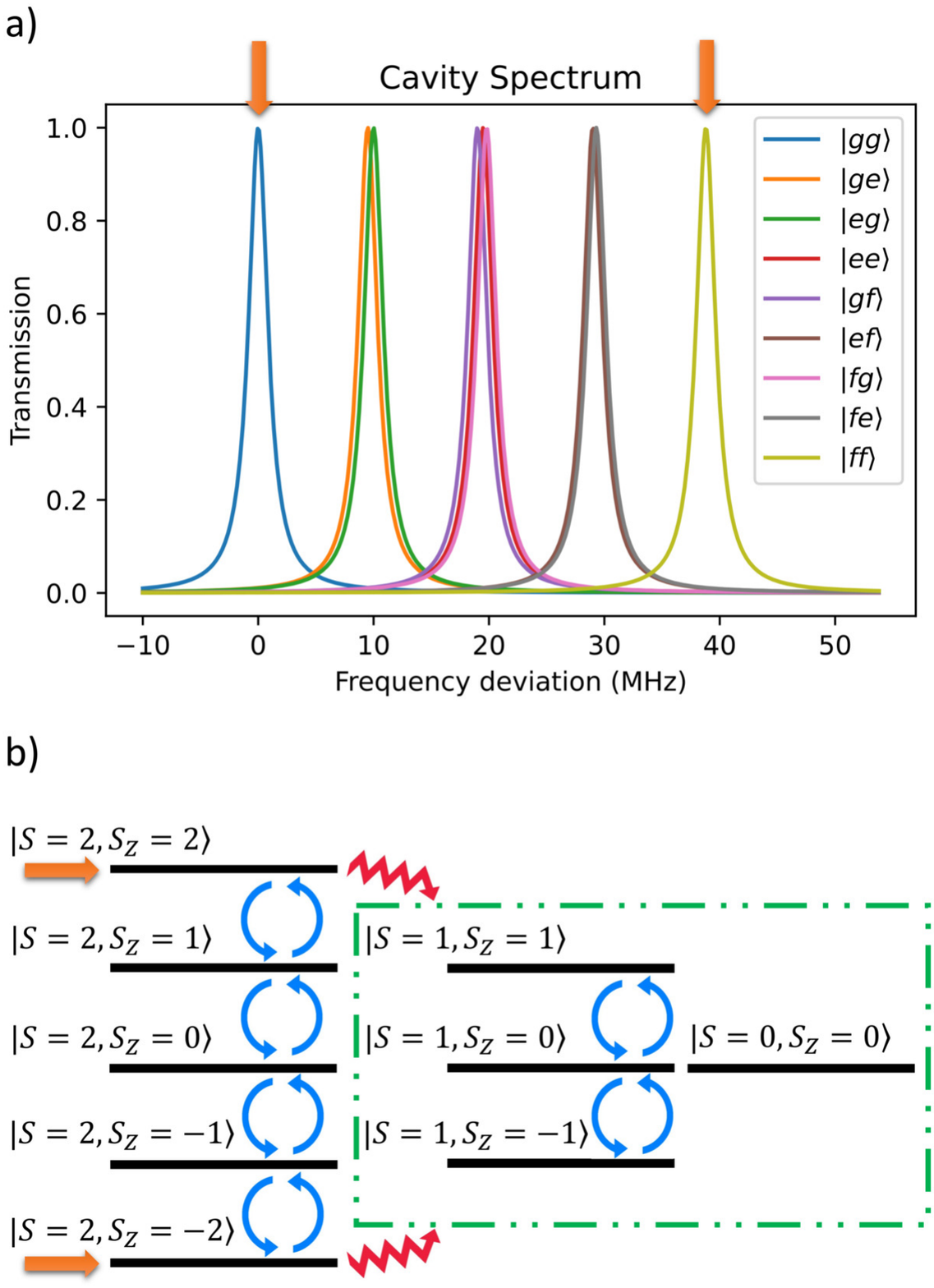}
\caption{\label{fig2II}  {\bf Diagram of the two-qutrit protocol}. (a) The cavity spectrum shifted by different two-qutrit states, with two cavity probes applied at the frequencies marked by the orange arrows. (b) Hilbert space engineering for the stabilization of two adjacent qutrits into the $S_\mathrm{total} \in \{0,1\}$ subspace. The orange arrows mark the states measured by the two cavity probes shown in (a), and the blue arrows represent the ``0-photon drives''. The green rectangle encircles the stabilized subspace of $S_\mathrm{total} \in \{0,1\}$, and the zigzag red arrows are dissipative processes steering into the stabilized subspace, with a similar mechanism as shown in the inset of Fig.~\ref{fig2I}(b). }  
\end{figure}

\subsection{Dissipative stabilization protocol} \label{section3B}

\kz{The AKLT Hamiltonian reviewed above} is frustration-free with a unique ground state on periodic boundary conditions and a four-fold degenerate ground state subspace with open boundary conditions.
Thus, the AKLT state can be reached by the strategy of driving each two adjacent spin-1 particles out of the $S_\mathrm{total}=2$ subspace, as is manifested by the projector form of $\hat{\mathrm{H}}_\mathrm{AKLT}$ in Eq.~\ref{eqn:AKLTproj}. 
Figure \ref{fig2II} presents an extension of the two-qubit protocol onto a two-qutrit system, which instead stabilizes the system into the two-qutrit subspace representing $S_\mathrm{total} \in \{0,1\}$.

Considering two qutrits coupled to a common linear cavity in the strong dispersive regime, without applied drives, the effective Hamiltonian becomes
\begin{equation}
\label{eqn:disp_qutrit}
\mathrm{\hat{H}_{eff}}=\hbar \sum_{j} 
(\chi_j^\mathrm{A} |j\rangle^\mathrm{A} \langle j|^\mathrm{A}+\chi_j^\mathrm{B} |j\rangle^\mathrm{B} \langle j|^\mathrm{B}) \hat{a}^\dag \hat{a},
\end{equation}
\refeo{in the rotating frame for qutrit levels and the cavity, with rotating wave approximation and then the dispersive limit applied.}
Here $|j\rangle^\mathrm{A(B)}$ are the energy eigenstates of qutrit A(B). $\hbar \chi_j^\mathrm{A(B)}$ are the interaction energies between the cavity and qutrit A(B), indicating the cavity resonance frequency's shift summed over A and B. The cavity's resonance frequencies are shown in Fig.~\ref{fig2II}(a).

Inspired by the qubit protocol, the cavity is driven at its resonance frequencies for qutrit states $|gg\rangle$ and $|ff\rangle$, which acts as part of the ``pump''. With cavity photon number $n$, the qutrit energy levels are shifted to 
$\hbar\omega_j^\mathrm{A(B)}+n\hbar\chi_j^\mathrm{A(B)}$.
Here, the three anharmonic energy levels are denoted as $|g\rangle$, $|e\rangle$ and $|f\rangle$, with $\omega_{ef}^\mathrm{A(B)}=\omega_f^\mathrm{A(B)}-\omega_e^\mathrm{A(B)}$, $\omega_{ge}^\mathrm{A(B)}=\omega_e^\mathrm{A(B)}-\omega_g^\mathrm{A(B)}$, $\chi_{ef}^\mathrm{A(B)}=\chi_f^\mathrm{A(B)}-\chi_e^\mathrm{A(B)}$,
$\chi_{ge}^\mathrm{A(B)}=\chi_e^\mathrm{A(B)}-\chi_g^\mathrm{A(B)}$,
and $\chi_{gf}^\mathrm{A(B)}=\chi_f^\mathrm{A(B)}-\chi_g^\mathrm{A(B)}$.  
Thus, we apply the ``0 photon drive'' at $\omega_{ge}^\mathrm{A(B)}$ and $\omega_{ef}^\mathrm{A(B)}$ simultaneously with the same Rabi frequency 
$\Omega^{(0)}$, while the ``$n$ photon drive'' is applied at $\omega_{ge}^\mathrm{A(B)}+n\chi_{ge}^\mathrm{A(B)}$ and $\omega_{ef}^\mathrm{A(B)}+n\chi_{ef}^\mathrm{A(B)}$   with Rabi frequency $\Omega^{(n)}$. 
\refeo{The effective Hamiltonians for those continuous drives are given by}
\begin{equation}
\begin{split}
&\hat{H}_\mathrm{eff}^{(0)}\propto\hat{S}_x^\mathrm{A} \otimes I^\mathrm{B}+I^\mathrm{A} \otimes \hat{S}_x^\mathrm{B},
\\
&\hat{H}_\mathrm{eff}^{(n)}\propto\hat{R}_{gf}^\mathrm{A} \otimes I^\mathrm{B}-I^\mathrm{A} \otimes \hat{R}_{gf}^\mathrm{B},
\label{eqn:rot}
\end{split}
\end{equation}
\refeo{where $\hat{S}_x^\mathrm{A(B)}$ are the spin angular momentum operators of the spin-1 particles represented by qutrit A(B), with $\hat{S}_x^\mathrm{A(B)}=|g\rangle\langle e| + |e\rangle\langle f|+h.c.$, and $\hat{R}_{gf}^\mathrm{A(B)} = \ket{g}\bra{f} + \mathrm{h.c.}$ The former is implemented with in-phase and equal amplitude independent Rabi drives between state $|g\rangle$ and $|e\rangle$ and between state $|e\rangle$ and state $|f\rangle$, and the latter is induced by direct two-photon transition on qutrit A(B).  The precise form of these drives is given in Appendix~\ref{app:c} and Appendix~\ref{app:d} we discuss alternative ``$n$ photon drives''.} 
The effective drive Hamiltonian $\hat{H}_\mathrm{eff}^0$ coincides with the total spin angular momentum $\hat{S}_x^\mathrm{total}$. 
Thus, this drive preserves the total spin represented by the two-qutrit system, while it rotates between different eigenstates of the $z$ component for the total spin $\hat{S}_z^\mathrm{total}$. 
 Meanwhile, the ``$n$-photon drive'' does not preserve the total spin and has non-zero components linking states $|gg\rangle|\bar{n}\rangle^\mathrm{C}$ and $|ff\rangle|\bar{n}\rangle^\mathrm{C}$ to the $|S_\mathrm{total}=0\rangle|\bar{n}\rangle^\mathrm{C}$ and $|S_\mathrm{total}=1\rangle|\bar{n}\rangle^\mathrm{C}$ subspace. 
Similar to the process described in Section \ref{section2B}, with those two drives combined, the two-qutrit system undergoes unidirectional evolution into the target subspace.
The autonomous feedback loop here provides quantum jump operators from $|gg\rangle$ and $|ff\rangle$ to $S_\mathrm{total} \in \{0,1\}$ subspace, with an overall rate proportional to the cavity linewidth $\kappa$.

Noticing that states $|gg\rangle$ and $|ff\rangle$ represent two spin-1 particles' states $|S^\mathrm{total}=2, S_z^\mathrm{total}=2\rangle$ and $|S^\mathrm{total}=2, S_z^\mathrm{total}=-2\rangle$, the drive $\hat{S}_x^\mathrm{total}$ thus connects the entire subspace $S_\mathrm{total}=2$ with $|gg\rangle$ and $|ff\rangle$.
This ``0-photon drive'', when applied continuously, therefore assists to evacuate the $S_\mathrm{total}=2$ subspace, leading to their stabilization into the target subspace where $S_\mathrm{total} \in \{0,1\}$. 
Meanwhile, the drive $\hat{S}_x^\mathrm{total}$ has no cross term between the $S_\mathrm{total}=2$ subspace and the target subspace, preventing leakage back from the stabilized states.
\refeo{Consequently, the applied drives and dissipations ensure that the degenerate AKLT manifold is a fixed point when the protocol is applied on a qutrit chain. Since the ``0 photon drive'' continues to induce rotations within this manifold, dynamics continue within the AKLT manifold once the qutrit chain has been stabilized.} 

 As is discussed in Section \ref{section2B}, the overlaps between the peaks in Fig.~\ref{fig2II}(a) can cause redundant measurements, compromising the fidelity of the target state.
In the stabilization protocol for the two-qubit Bell state, the cavity-qubit detuning can be tuned in order to achieve consistency between cavity shifts \cite{Shankar2013}.
However, when it comes to the two-qutrit protocol, where the stabilized subspace is four-dimensional, such discrepancies are unavoidable even with full tunability on the device parameters.
Thus, finite $\chi/\kappa$ becomes one of the main limiting factors for the final fidelity. Possible optimization paths are discussed in Appendices B and E.

\subsection{Numerical simulations}
\label{section3C}
Following the protocol discussed in Section \ref{section3B}, we model the qutrits and cavities in Qutip \cite{Johansson2012,Johansson2013} for numerical simulations, with the microwave drives applied during the stabilization process. Details of the simulation are given in Appendix~\ref{app:c}. Analyzing the simulation results, we now study the protocol's effectiveness as well as its performance with an increased number of qutrits in the chain.  

\begin{figure}
\centering
\includegraphics[width = 0.475\textwidth]{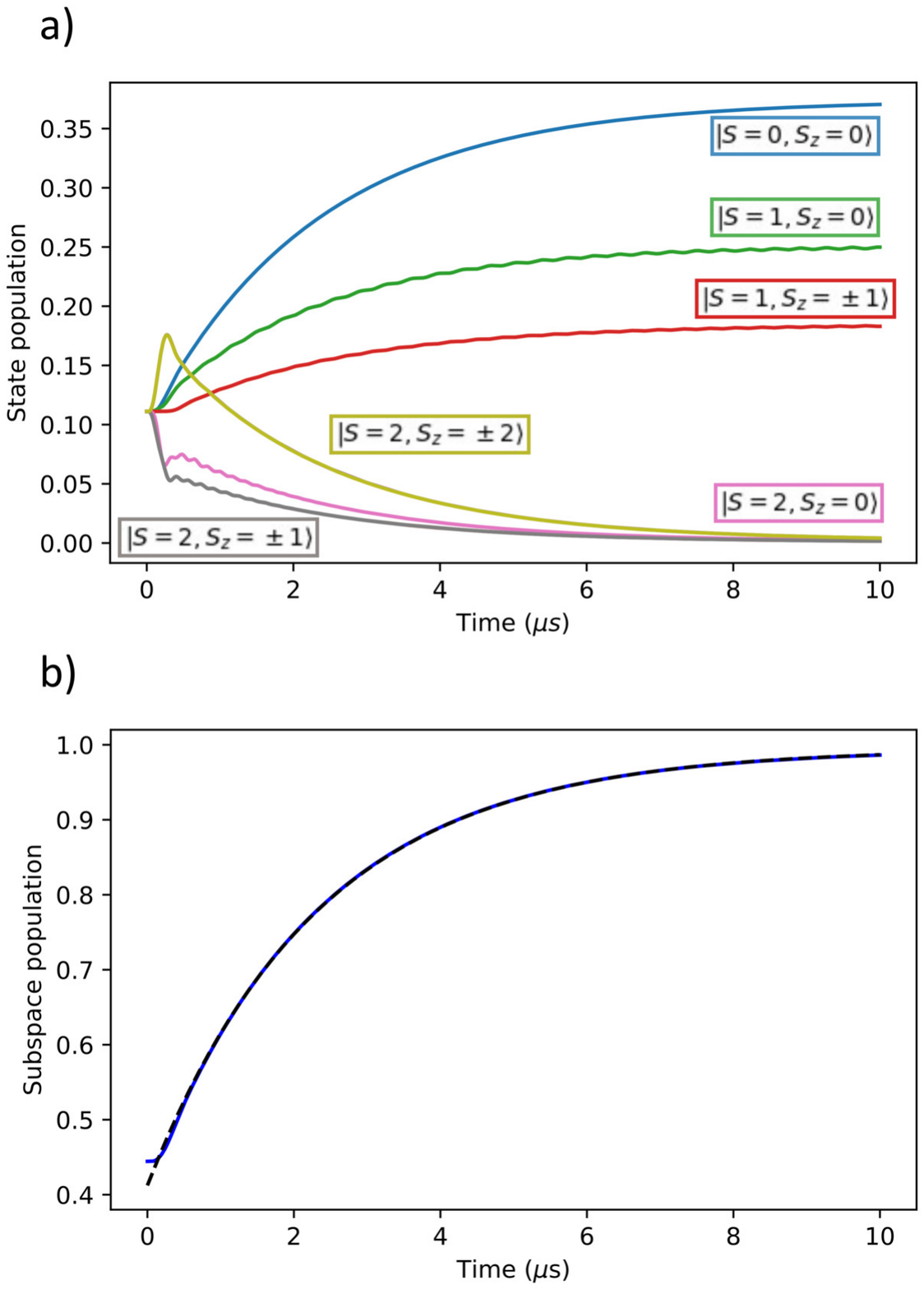}

\caption{\label{fig3}  {\bf Time evolution of two qutrits under the stabilization protocol}. (a) Stabilization process for one pair of neighboring qutrits, with the entire subspace $S_\mathrm{total} = 2$ eliminated and subspace $S_\mathrm{total} \in \{0,1\}$ stabilized. The colored lines are the \kz{the relative weights of the states, given by the diagonal entries of the density matrix} for eigenstates of both $\hat{S}_\mathrm{total}$ and $\hat{S}_\mathrm{total}^z$. \refeo{Although the specific distribution of states inside the AKLT depends on the initial state distribution as well as our choice of ``$n$-photon drive'', here we are just concerned about stabilizing into the entire subspace.} (b) The blue line represents the change of total population in the targeted subspace where $S_\mathrm{total} \in \{0,1\}$. The black dotted line is an exponential fit. }
\end{figure}

As is shown in Fig.~\ref{fig3}(a), an adjacent pair of qutrits is initialized in a maximally mixed state of the nine-dimensional Hilbert space. \kz{This choice of initial state is only a matter of  convenience and not necessary for the protocol.}  The system then evolves under the driven dissipative protocol which consists of always-on drives. For the qutrit pair, all five states representing $S_\mathrm{total} =2$ have their state population converging to zero in the course of the protocol, while the four states in subspace $S_\mathrm{total} \in \{0,1\}$ are preserved and stabilized.
The protocol effectively eliminates the $S_\mathrm{total} =2$ subspace while steering the system into the $S_\mathrm{total} \in \{0,1\}$ subspace. 
Figure \ref{fig3}(b) shows the total four-state population in subspace $S_\mathrm{total} \in \{0,1\}$, which is the two-qutrit AKLT subspace. \refeo{The curve can be well fitted with an exponential function $y=Ae^{-bx}+C$.} In later analysis, the fitting parameter $C$ is extracted as the final fidelity of the target subspace and $b$ as the stabilization rate, with the convergence time for the protocol calculated as $1/b$.

The protocol, therefore, drives the spin-1 chain into the AKLT manifold. The resulting quantum state within this manifold may be a pure state, contain coherences within the AKLT manifold, or be a mixed state depending on the initial conditions at the start of the protocol. 

\begin{figure}
\centering
\includegraphics[width = 0.5\textwidth]{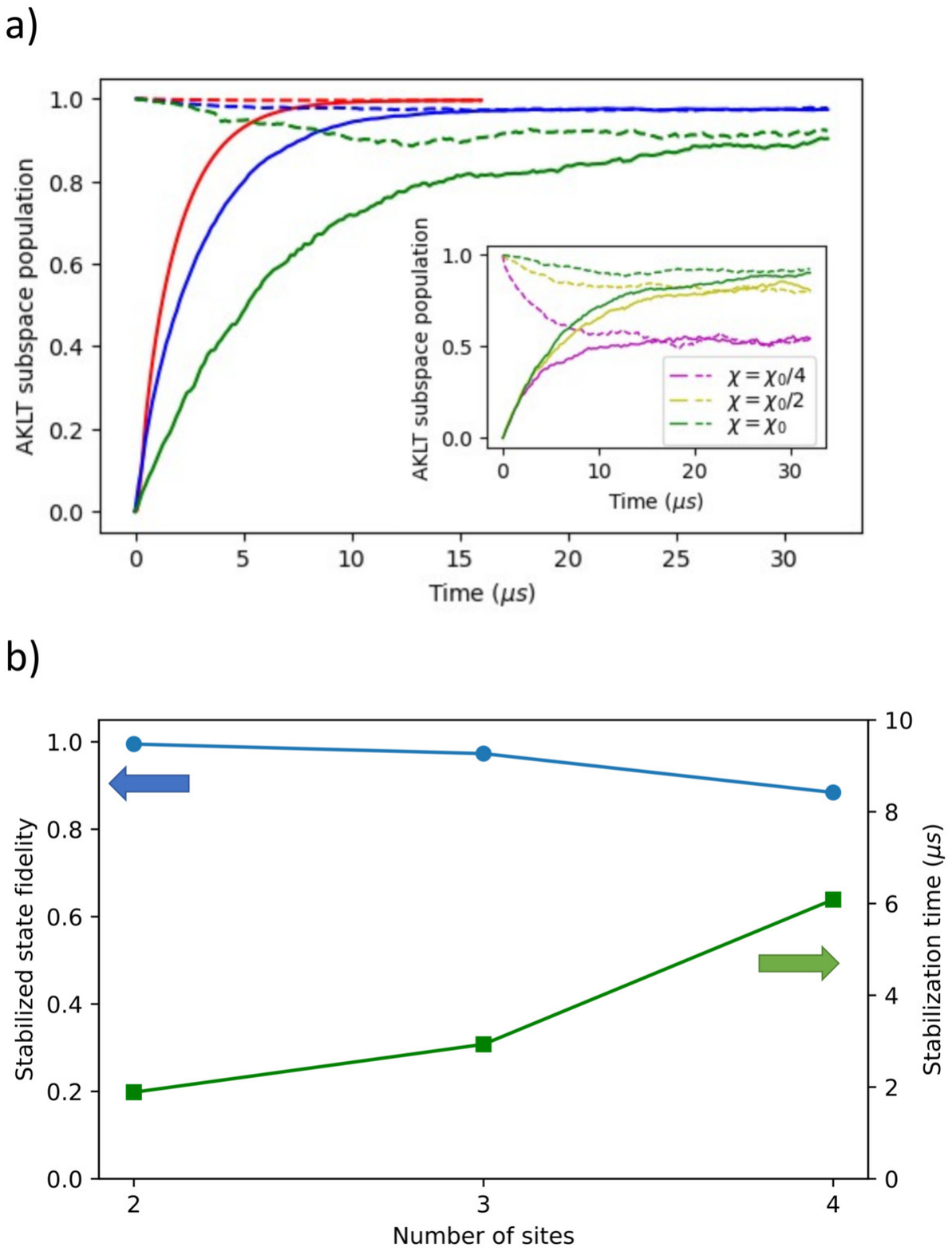}

\caption{\label{fig4}  {\bf Protocol performance for larger qutrit chains}. 
(a) Starting from the qutrits' ground state $\ket{g\ldots g}$ (solid colored lines) and starting within the AKLT subspace (dashed colored lines), we simulate the time evolution of the \kz{AKLT chain with open boundary conditions} subspace population under the driven dissipative protocol with \refeo{varying number of qutrits,} $N_\mathrm{sites} = 2$ (red), $N_\mathrm{sites} = 3$ (blue) and $N_\mathrm{sites} = 4$ (green) for comparison. Inset: The four qutrit time evolution under the stabilizing protocol in terms of the system population in the AKLT subspace, with the cavity shift scaling as $\chi_0$, $\chi_0/2$, and $\chi_0/4$.
(b) Extracted fitting parameters as the final fidelity and the convergence time \refeo{with the same method as in Fig.~\ref{fig3} (b) on the total population in the AKLT subspace}. The blue dots represent the varied final fidelity with systems of two, three, and four qutrits (left axis), and the green dots represent the convergence time for the protocol (right axis). }
\end{figure}

Extended from the two-qutrit case, we study the evolution of the system with 2, 3, and 4 \refeo{qutrits} in the AKLT chain. \refeo{Here, the effective Hamiltonian of the ``0-photon drive'' and the ``$n$-photon drive'' can be viewed as}
\begin{equation}
\begin{split}
&\hat{H}_\mathrm{eff}^{(0)}\propto\sum_{i=1}^{N_\mathrm{sites}} \hat{S}_x^i  ,
\\
&\hat{H}_\mathrm{eff}^{(n),i}=\hat{R}_{gf}^i - \hat{R}_{gf}^{i+1}.
\end{split}
\end{equation}
\refeo{The ``0-photon drive'' is applied by in-phase, equal amplitude single qutrit $S_x$ drive on each site. The ``$n$-photon drives'' are carried out with the $i_{th}$ pair of qutrits, qutrit $i$ and qutrit $i+1$, at the shifted qutrit frequencies due to the $i_{th}$ cavity. We notice that the global drive $\hat{H}_\mathrm{eff}^{(0)}$ commutes with the AKLT Hamiltonian. So, once the system is in the AKLT subspace, applying the ``0-photon drive'' will not rotate the system out of the ground state manifold. In contrast, the ``$n$-photon drives'' are defined by each individual resonator and come into effect whenever there is an adjacent bond with $S_\mathrm{total}=2$. }

We consider two initial preparations: either one of the  AKLT states or the product state of single-qutrit ground states, $|g\ldots g\rangle$. 
\refeo{For the choice of the former, since we observe that the decay features of the AKLT subspace total population are similar for different initial states within the subspace, we simply initialize the simulation with a particular state within the AKLT subspace.}
The evolution of the populations in the AKLT subspace under the protocol is shown in Fig.~\ref{fig4}(a). 
Here we can extract the final fidelity and stabilization time for different chain sizes \refeo{defined by the same method as in Fig.~\ref{fig3}(b)}, which is presented in Fig.~\ref{fig4}(b).
While stabilization of the AKLT state is observed, the final fidelity decreases for larger chains. 

\begin{figure}
\centering
\includegraphics[width = 0.5\textwidth]{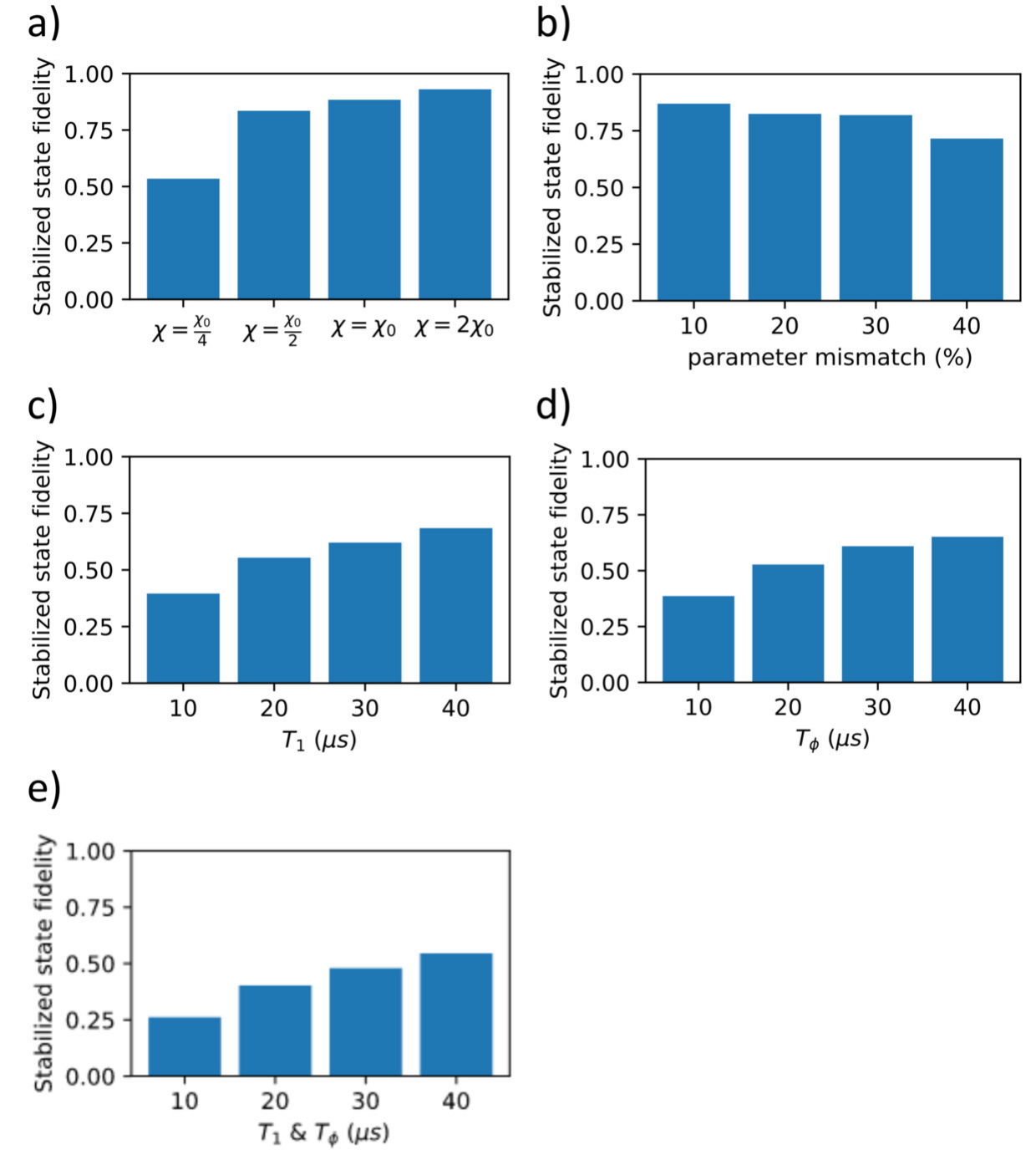}
\caption{\label{fig5}  {\bf Fidelity in the four-qutrit AKLT subspace estimated under experimental imperfections}. \refeo{The stabilized state fidelity is defined by the same method as in Fig.~\ref{fig3} (b) according to the tendency of the state population vs. time.} (a) The impact of finite dispersive shift $\chi$ (Sec.~\ref{sec2a}) to cavity decay rate $\kappa$ (Sec.~\ref{section2B}), smaller $\chi/\kappa$ degrades the stabilized state fidelity. (b) Running the protocol with mismatched device parameters. The protocol is relatively robust to small parameter mismatches. (c,d,e) Evaluating the protocol fidelity under the dissipation channel with finite qutrit $T_1$/$T_\phi$/$T_1 \& T_\phi$ between adjacent levels. The final state fidelity is degraded when the qutrit lifetime becomes comparable with the stabilization time.}
\end{figure}

The decrease in final fidelity is due to the finite value of $\chi/\kappa$, which we explore in the inset of Fig.~\ref{fig4}(a).  
Here we plot the AKLT state population starting from the AKLT state or the ground state with cavity shifts $\chi$ given as $\chi_0$, $\chi_0/2$, and $\chi_0/4$. 
\refeo{As is shown in inset of Fig.~\ref{fig4}(a) and in Fig.~\ref{fig5}(a), smaller values of $\chi$ bring extra measurements on the preserved two-qutrit states, causing dephasing out of the AKLT state. This effect of extra dephasing is similar to the effect caused by a shortened $T_2$, reducing the fidelity as $N_\mathrm{sites}$ increases since the extra dephasing occurs for each pair of neighboring qutrits. 
For similar reasons, the stabilization time increases for larger qutrit chains because these errors can diffuse around the chain before being eliminated by the protocol. }

\refet{For this reason, the protocol favors a large value of $\chi$, however, reasonable values of $\chi$ are an experimental limitation set by the dispersive condition. 
The value of $\chi_0$ we choose here is quite large for typical transmon systems. We have chosen this value for clarity of demonstration. However, as we explore in Appendix~\ref{app:e}, large effective values of $\chi$ can be achieved with very reasonable experimental parameters. Such a realistic implementation requires a more complex set of drives, yet achieves even higher state fidelities than the symmetric case presented in the main text.}


In actual experiments, the qutrits are not perfectly realizable as in the model simulated above. For example, there is limited control over the qutrit-cavity coupling parameters.
In the above simulations, we assume that the cavity frequency shifts induced by qutrits are approximately equivalent, with discrepancies smaller than 5 percent. 
This relation requires equally spaced cavity frequency shifts from qutrit states $|g\rangle$, $|e\rangle$, and  $|f\rangle$, as well as equal cavity-qutrit interaction terms among different qutrits. 
Such an assumption is made for the sake of simplicity, but may be  difficult to realize in experiments. 

In the experimental realization with transmon qubits \cite{Shankar2013} matching between the cavity shift of qubit A and B, $\chi_\mathrm{A}=\chi_\mathrm{B}$, was achieved by tuning the qubit frequencies, where $\chi_\mathrm{{A(B)}}\propto g_\mathrm{{A(B)}}^2/(\omega_\mathrm{{A(B)}}-\omega_r)$ with $g$ denoting the coupling strength.   
However, for the current protocol involving qutrits, matching all the dispersive shifts by simple frequency tuning of the qutrit levels is not possible.
Figure \ref{fig5}(b) shows the protocol fidelity on a four-qutrit chain with qutrit-cavity coupling parameters mismatched to varied degrees; The choice of device parameters is introduced in Appendix~\ref{app:b}. The final fidelity of the protocol decreases slowly from $\sim 85\%$ to $\sim 70 \%$ with larger parameter mismatches from $10\%$ to $40\%$, indicating a relatively small impact on the protocol performance. 

\refeo{Overall, the protocol retains a robustness to parameter mismatch because the  ``0-photon drive'' can always be achieved by adjusting the amplitude and phase of single qutrit drives regardless of the qutrit parameters. Also, between each two adjacent qutrits, the consistency of cavity shifts from $|g\rangle$ to $|f\rangle$ can be always achieved with one-by-one tuning of qubit frequencies, allowing the application of the ``$n$-photon drive''. These tunings allow variations in qutrit parameters in the array can be tolerated. Also, there are no specific requirements for the relation between cavity shifts of a single qutrit from state $|g\rangle$ to $|e\rangle$ or from state $|g\rangle$ to $|f\rangle$, to aim for a good protocol fidelity. The slight decrease in the fidelity results from increased extra dephasing induced by residual measurements, since the spectrum peaks shown in Fig.~\ref{fig2II} (a) no longer perfectly overlap considering the parameter mismatch. This effect can nevertheless be eliminated by an even higher ratio of $\chi/\kappa$, as is described in Appendix~\ref{app:b}}  

The other aspect of imperfect qutrits considers their intrinsic relaxation and dephasing.
In previous numerical simulations, we set the qutrit $T_1$ and $T_2$ at an optimistically high level ($T_1=T_2=500\ \mu$s, \refet{thus $T_\phi = 1000\ \mu$s}) for isolating the protocol performance from the effects brought up by extra environmental coupling. 
However, in realistic setups, the relaxation towards the ground state of every single qutrit, as well as the decoherence between qutrit levels, will drive the manybody system out of the AKLT subspace, impairing the effectiveness of the protocol. Nevertheless, we find that the driven dissipative protocol is still able to stabilize the system coherence far beyond the single qutrit coherence time. As is shown in Fig.~\ref{fig5}(c,d,e), the final fidelity in the AKLT subspace is extracted for finite values of $T_1$ and $T_{\phi}$, \refeo{as well as both $T_1$ and $T_\phi$, between adjacent qutrit levels. In Fig.~\ref{fig5}(c,d) we consider the case where we keep $T_1$ at the optimistically high value when we set $T_\phi$ to a limited value, and vice versa. With $T_1$ or $T_{\phi}$ solely set to a finite value from $10~\mu$s to $40~\mu$s, the protocol final fidelity increases from $\sim 40\% $ to $\sim 70\% $.
With $T_1$ and $T_\phi$ both set to a finite value from $10~\mu$s to $40~\mu$s [Fig.~\ref{fig5}(e)], the AKLT subspace fidelity increases from $\sim 25\% $ to $\sim 55\% $.} 
\refeo{Considering the $3^4=81$ dimensions of the four-qutrit Hilbert space, where a maximally mixed state would have a $\sim5$\% fidelity to the ALKT subspace, this result further confirms the ability of a driven-dissipative method to maintain 
the state coherence beyond native} relaxation or dephasing time.

\section{discussion}

In this work, we proposed and analyzed a driven-dissipative protocol to achieve preparation and stabilization of the AKLT state \refeo{with OBC} in a one-dimensional superconducting qutrit array.
\refeo{Our stabilization of a manifold of edge states opens up  the opportunity to study dynamics within that manifold, rather than just preparing a pure state.}
Via numerical simulation, we verified the effectiveness of our protocol for a pair of two adjacent qutrits as well as with extended system sizes. 

Our results illuminate the possibility for efficient generation of manybody entangled states on superconducting qutrit platforms. Generalization from the AKLT state to other matrix product states or projected entangled pair states \cite{Cirac2021} may be possible. Since the spin-1 ALKT state represents a quantum wire in measurement-based quantum computing, it only allows for performing quantum computation of a limited scope.  Universal computation, in contrast, can be realized by the spin-2 AKLT state on a 2-dimensional square lattice  \cite{Wei2015}, which would be a natural next step beyond our work. 

\section{Acknowledgements} 

\kz{We thank A. Seidel for helpful comments.} This research was supported by NSF Grant No. PHY-1752844 (CAREER) and the Air Force Office of Scientific Research (AFOSR) Multidisciplinary University Research Initiative (MURI) Award on Programmable systems with non-Hermitian quantum dynamics (Grant No. FA9550-21-1- 0202). K.S. acknowledges support by Deutsche Forschungsgemeinschaft (DFG, German Research Foundation) grant No. GO 1405/6-1.  \kz{Y.G. acknowledges  DFG-CRC grant No. 277101999 - TRR 183 (Project C01),  DFG grants No. EG
96/13-1, the Helmholtz International Fellow Award, and the Israel Binational Science Foundation--National Science Foundation through award DMR-2037654.}

\appendix


\section{The dispersive limit} \label{app:a}

\refet{Here we provide a detailed derivation of the system effective Hamiltonian in the dispersive limit, as in Eqn.~\ref{eqn:disp_qubit} and Eqn.~\ref{eqn:disp_qutrit} (equivalent to Eqn.~\ref{eqn:disp_qutrit_simu}).  
For two transmon qudits coupled to a common linear cavity \cite{Blais2004,Wallraff2004}, after applying the rotating wave approximation, the system Hamiltonian should be,}
 \begin{equation}
     \begin{split}
     &\hat{H} = \hat{H}_0 + \hat{V},
     \\
     &\hat{H}_0 = \hbar \sum_j \omega_j^A |j\rangle^A \langle j|^A + \hbar \sum_j \omega_j^B |j\rangle^B \langle j|^B  + \hbar \omega_r \hat{a}^\dagger \hat{a},
     \\
     &\hat{V} = \hbar \sum_i g_{i,i+1}^A (|i\rangle^A \langle i+1|^A \hat{a}^\dagger+|i+1\rangle^A\langle i|^A \hat{a} )
     \\
     &+\hbar \sum_i g_{i,i+1}^B (|i\rangle^B\langle i+1|^B \hat{a}^\dagger+|i+1\rangle^B\langle i|^B \hat{a} ).
     \end{split}
\end{equation}
\refet{Here the coupling strength $g_{i,i+1}\approx \sqrt{i+1}g_0$ \cite{Koch2007}.
In the dispersive limit, the detuning between adjacent energy level differences of the qutrit is much larger than the coupling strengths. This allows us to perform the Schrieffer-Wolff transformation, which can approximately diagonalize the system Hamiltonian in the dispersive limit, $\hat{H}_\mathrm{eff} = e^{-\hat{S}} \hat{H} e^{\hat{S}} $, with }
\begin{equation}
\begin{split}
&\hat{S}= \sum_i\lambda^A_i (|i+1\rangle^A\langle i|^A \hat{a}-|i\rangle^A\langle i+1|^A \hat{a}^\dagger )
\\
&+\sum_i\lambda^B_i (|i+1\rangle^B\langle i|^B \hat{a}-|i\rangle^B\langle i+1|^B \hat{a}^\dagger ),
\end{split}
\end{equation} 
where $\lambda_i^{A(B)} = g^{A(B)}_{i,i+1}/(\omega^{A(B)}_{i,i+1}-\omega_r)$, and $\omega^{A(B)}_{i,i+1}=\omega^{A(B)}_{i+1}-\omega^{A(B)}_{i}$. 
We can have
\begin{equation}
\begin{split}
&[\hat{S},\hat{H}_0]= \hbar \sum_i \lambda_i^A (-\omega^A_{i+1}+\omega^A_i + \omega_r) (|i+1\rangle^A \langle i |^A \hat{a} 
\\
&+ |i\rangle^A \langle i+1|^A \hat{a}^\dagger)+\hbar \sum_i \lambda_i^B (-\omega^B_{i+1}+\omega^B_i+ \omega_r)
\\
& (|i+1\rangle^B\langle i |^B \hat{a} + |i\rangle^B \langle i+1|^B \hat{a}^\dagger),
\\
&[\hat{S},\hat{V}]=\hbar \sum_i (\lambda^A_{i+1}g^A_{i,i+1} - \lambda_i g^A_{i+1,i+2} )(|i+2\rangle^A \langle i|^A \hat{a}\hat{a}
\\
&+|i\rangle^A \langle i+2 |^A \hat{a}^\dagger \hat{a}^\dagger )+ 2 \hbar \sum_i \chi^A_{i,i+1} |i+1\rangle^A \langle i+1|^A+
\\
&  2 \hbar \sum_{i=1}^\infty (\chi^A_{i-1,i}-\chi^A_{i,i+1}) |i\rangle^A \langle i |^A \hat{a}^\dagger \hat{a}
-2 \hbar \chi_{0,1}^A |0\rangle^A \langle 0|^A \hat{a}^\dagger \hat{a}
\\
&+\hbar \sum_i (\lambda^B_{i+1}g^B_{i,i+1} - \lambda_i g^B_{i+1,i+2} )(|i+2\rangle^B \langle i|^B \hat{a}\hat{a}
\\
&+|i\rangle^B \langle i+2 |^B \hat{a}^\dagger \hat{a}^\dagger )+ 2 \hbar \sum_i \chi^B_{i,i+1} |i+1\rangle^B \langle i+1|^B+
\\
&  2 \hbar \sum_{i=1}^\infty (\chi^B_{i-1,i}-\chi^B_{i,i+1}) |i\rangle^B \langle i |^B \hat{a}^\dagger \hat{a}
-2 \hbar \chi_{0,1}^B |0\rangle^B \langle 0|^B \hat{a}^\dagger \hat{a}
\\
&+\hbar(\sum_i g_{i,i+1}^B |i\rangle^B\langle i+1|^B)(\sum_i \lambda_i^A |i+1\rangle^A\langle i|^A )
\\
&+\hbar(\sum_i g_{i,i+1}^B |i+1\rangle^B\langle i|^B)(\sum_i \lambda_i^A |i\rangle^A\langle i+1|^A )
\\
&+\hbar(\sum_i \lambda_i^B |i+1\rangle^B\langle i|^B )(\sum_i g_{i,i+1}^A |i\rangle^A\langle i+1|^A)
\\
&+\hbar(\sum_i \lambda_i^B |i\rangle^B\langle i+1|^B )(\sum_i g_{i,i+1}^A |i+1\rangle^A\langle i|^A),
 \end{split}
\end{equation}
\refet{where the relation $[\hat{S},\hat{H}_0]+\hat{V}=0$ stands.
Also, from the Baker-Campbell-Hausdorff relation, there is $\hat{H}_\mathrm{eff} = \hat{H}_0 + \hat{V} + [\hat{S},\hat{H}_0]+[\hat{S},\hat{V}] + \frac{1}{2} [\hat{S},[\hat{S},\hat{H}_0]]+\frac{1}{2}[\hat{S},[\hat{S},\hat{V}]]+ \ldots$. 
Then we have, 
$\hat{H}_\mathrm{eff} = \hat{H}_0 + \frac{1}{2} [\hat{S},\hat{V}] + O(\lambda^2) $.
Here, the two-photon transition terms involving $\hat{a}\hat{a}$ and $\hat{a}^\dagger \hat{a}^\dagger$ are small and can be omitted \cite{Koch2007}. We also notice that cavity-mediated interaction terms emerge between next-nearest neighbor layout of qutrits. However, when the qutrits in the array are also far-detuned from each other, this term becomes counter-rotating when we go to the rotating frame of the qutrit and the cavity. The final Hamiltonian, in the rotating frame for the shifted qutrit Hamiltonian and the cavity frequency, becomes,}
\begin{equation}
H_\mathrm{eff}=
\hbar \sum_{i=0}^\infty (\chi_i^A |i\rangle^A \langle i |^A+\chi_i^B |i\rangle^B \langle i |^B) \hat{a}^\dagger \hat{a}.
\end{equation}
\refet{Here $\chi_i=(\chi_{i-1,i}-\chi_{i,i+1})$ for $i\geq1$ and $\chi_i=-\chi_{0,1}$ for $i=0$, and $\chi_{i,i+1}$ is defined as $\lambda_i g_{i,i+1}$ which is $g_{i,i+1}^2 /(\omega_{i,i+1}-\omega_r)$. Then we obtain Eqn.~\ref{eqn:disp_qutrit}, which becomes Eqn.~\ref{eqn:disp_qutrit_simu} when we only consider the first three levels of the qudit. This becomes Eqn.~\ref{eqn:disp_qubit} when we only consider the first two levels. }

\section{The strong dispersive limit} \label{app:b}

In the strong dispersive limit, we have the relation $\chi \gg \kappa$ for the cavity-qutrit coupling parameters.
The cavity resonance amplitude, $T=1/(1+x^2)$, is of a Lorentzian spectral line shape, which is presented in Fig.~\ref{fig2I}(a) as well as in Fig.~\ref{fig2II}(a).
Here $x=2(\omega-\omega_r)/\kappa$ and $\omega_r$ is the cavity resonance frequency,
Thus, as we probe the system at one of the peaks, there is a high ratio between the resonance amplitudes for probed and unprobed states.
However, with a relatively long protocol time, a small resonance amplitude still causes residual measurements that distinguish between stabilized two-qutrit states. 
Such extra measurements induced by the probe, as introduced in Section~\ref{section3B}, have a visible effect on the final fidelity of the stabilization for the manybody entangled state. 
As is shown in the inset of Figure~\ref{fig4}(a) as well as Figure~\ref{fig5}(a), such an effect can hinder the scalability of our protocol if uneliminated. To explore this further, we study the performance of the protocol for smaller values of $\chi$, which exacerbates the residual measurement effect. Here the ``0 photon drives'' and the measurement probe are applied to systems with qutrit number $N_\mathrm{site}=3$ and $N_\mathrm{site}=4$, as is described in Section~\ref{section3B}. The system is initialized in one of the open boundary AKLT states, and the decreases in the AKLT subspace population are fitted with exponential functions $y=Ae^{-bx}+C$. The extracted dephasing rates $b$ are plotted in the insets of Figure~\ref{figs1}, showing 
that the decay rate grows significantly as $\chi$ is decreased.  
%
This trend favors larger values of $\chi$, which can be achieved by optimizing the device parameters, via enhanced coupling parameter $g$ or reduced cavity-qutrit frequency detuning, as well as by working in the so-called straddling regime for transmon circuits \cite{Koch2007} \refeo{as discussed in Appendix~\ref{app:e}}. On current platforms, we can expect efficient stabilization as long as the residual measurement-induced dephasing is reduced to some negligible level compared to the intrinsic dephasing and relaxation of the superconducting qutrits. 

\renewcommand{\thefigure}{S1}
\begin{figure}
\centering
\includegraphics[width = 0.475 \textwidth]{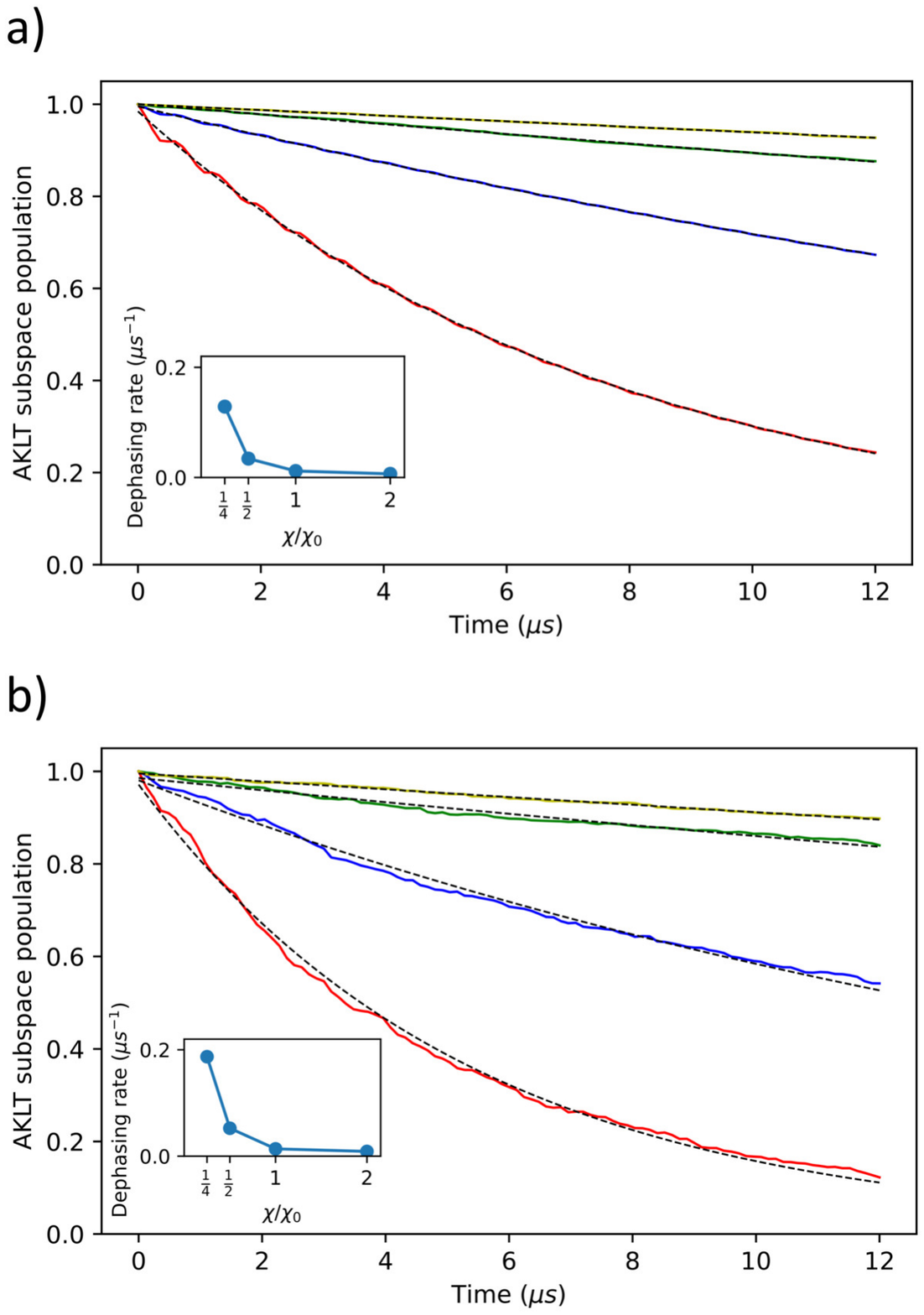}

\caption{\label{figs1}  {\bf Dephasing out of the AKLT subspace due to extra measurements caused by the cavity probes}. Initialized with a three- (panel a) or four- (panel b) qutrit AKLT state, the system dephases under the ``0-photon drives'' and the cavity probes, with qutrit induced cavity shift set as $\chi=\chi_0/4$ (red), $\chi=\chi_0/2$ (blue), $\chi=\chi_0$ (green) and $\chi=2\chi_0$ (yellow). The AKLT subspace population curves are fitted with exponential functions. Insets display the extracted dephasing rate of the system versus the relative cavity shift $\chi/\chi_0$.}
\end{figure}

\renewcommand{\thefigure}{S2}
\begin{figure}
\centering
\includegraphics[width = 0.475\textwidth]{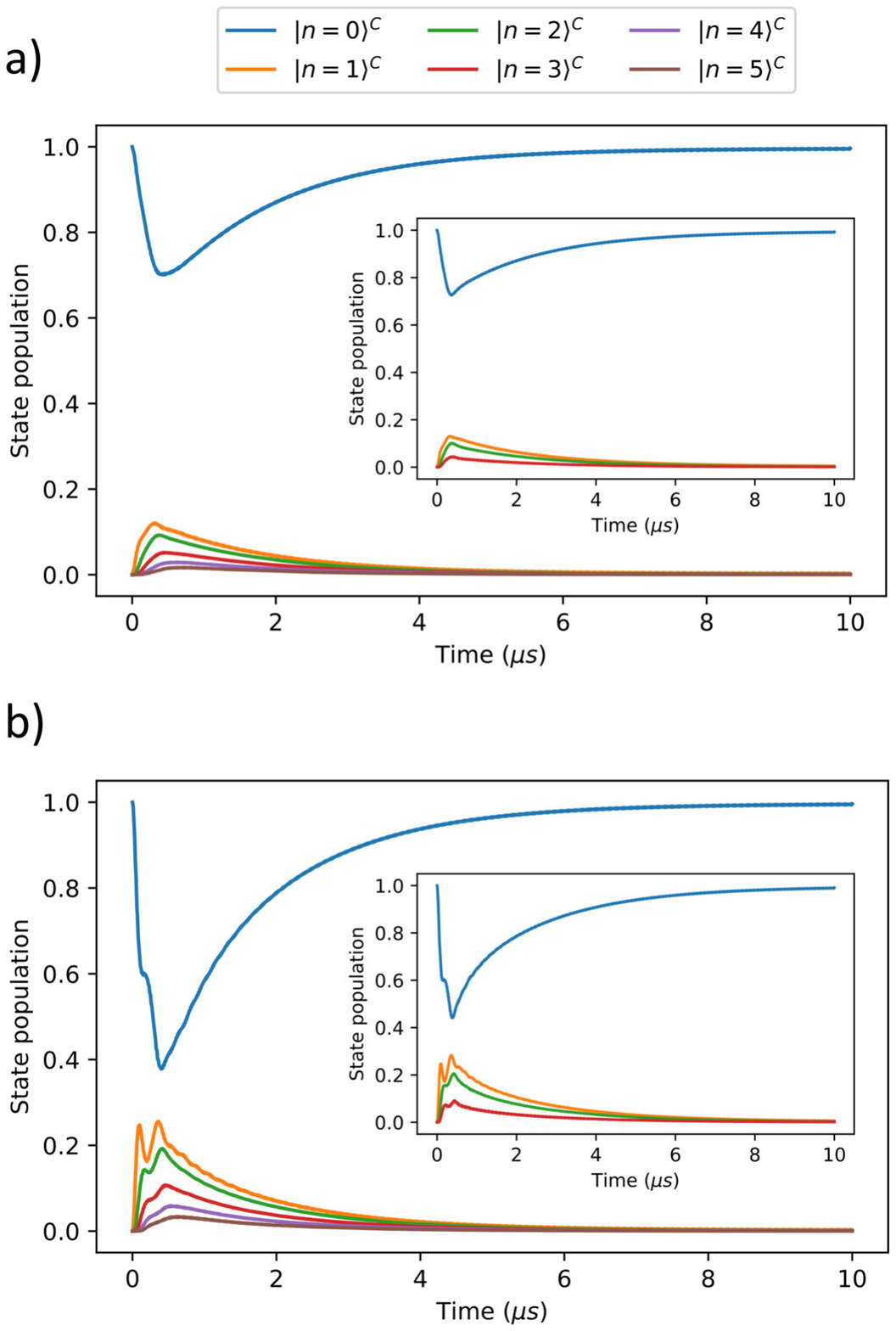}

\caption{\label{figs2}  {\bf Cavity photon population during the stabilization process}. (a) The system is initialized with a fully mixed state and then evolves under the protocol. (b)  The system is initialized in the ground state. Insets show the stabilization process calculated with a maximal photon number cutoff of $n=3$ instead of $n=9$. }
\end{figure}

\section{Numerical simulation setups} \label{app:c}

For the two-qutrit protocol, we simulate the Lindblad master equation

\begin{equation}
\begin{split}
&\frac{d}{dt}\rho(t)  =-\frac{i}{\hbar}[\mathrm{\hat{H}}(t),\rho (t)]+\kappa\mathbf{D}[\hat{a}]\rho (t) +
\\
&\sum_{\substack{ l=ge,ef \\ j=A,B }} \left ( \frac{1}{T_1^{j,l}} \mathbf{D} [\sigma_{-}^{j,l}] \rho (t) + \frac{1}{ 2 T_\phi^{j,l}} \mathbf{D} [\sigma_{z}^{j,l}] \rho (t) \right ).
\end{split}
\end{equation}

Here, $T_1^{A(B),ge}$ and $T_1^{A(B),ef}$ are the realaxation time from state $|e\rangle$ to state $|g\rangle$ and from state $|f\rangle$ to state $|e\rangle$. The pure dephasing rate is given by,
$$1/T_{\phi}^{A(B),ge(ef)}=1/T_2^{A(B),ge(ef)}-1/2T_1^{A(B),ge(ef)},$$
where $T_2^{A(B),ge(ef)}$ are the dephasing times between the corresponding two adjacent levels.
The Lindblad superoperator for an observable $\hat{O}$ acting on the density matrix $\rho$ is defined as

\begin{equation}
\mathbf{D}[\hat{O}]\rho = \hat{O} \rho \hat{O}^{\dagger} - \frac{1}{2} \hat{O}^{\dagger} \hat{O} \rho - \frac{1}{2} \rho \hat{O}^{\dagger} \hat{O}. 
\end{equation}
For the unitary part of system evolution, we have the Hamiltonian,
\begin{equation}
\mathrm{\hat{H}}=\mathrm{\hat{H}}_\mathrm{system}+\mathrm{\hat{H}}_\mathrm{probe}+\mathrm{\hat{H}}_0+\mathrm{\hat{H}}_n.
\end{equation}
Since we work in the rotating frame for the qutrit transition energies as well as for the center of the cavity resonance frequencies of corresponding to $|gg\rangle$ and $|ff\rangle$, this Hamiltonian consists of
\begin{equation}
\label{eqn:disp_qutrit_simu}
\begin{split}
\mathrm{\hat{H}}&_\mathrm{system}  = \hbar \left( \frac{\chi_{gf}^A}{2}\sigma_z^{A,gf} + \frac{\chi_{ge}^A-\chi_{ef}^A}{2}|e\rangle^A\langle e |^A \right. \\
& \left. +\frac{\chi_{gf}^B}{2}\sigma_z^{B,gf} + \frac{\chi_{ge}^B-\chi_{ef}^B}{2}|e\rangle^B\langle e |^B \right) \hat{a} ^ \dagger \hat{a},
\end{split}
\end{equation}
\begin{equation}
\begin{split}
\mathrm{\hat{H}}_\mathrm{probe}=2\hbar\epsilon_C \cos\left(\frac{\chi^A_{gf}+\chi^B_{gf}}{2} t \right)\left(\hat{a}+\hat{a}^\dagger\right),
\end{split}
\end{equation}
\begin{equation}
\begin{split}
\mathrm{\hat{H}}_0=\hbar\Omega^{(0)} \left(\sigma_x^{A,ge}+\sigma_x^{A,ef}+\sigma_x^{B,ge}+\sigma_x^{B,ef}\right),
\end{split}
\end{equation}
and,
\begin{equation}
\begin{split}
&\mathrm{\hat{H}}_n  =\hbar\Omega^{(n)} \left(\cos\left(n\frac{\chi^A_{gf}+\chi^B_{gf}}{2}t\right)(\sigma_x^{A,gf}-\sigma_x^{B,gf}) \right.\\
&\left.-\sin \left(n\frac{\chi^A_{gf}+\chi^B_{gf}}{2}t\right)(\sigma_y^{A,gf}-\sigma_y^{B,gf})\right).
\label{eqn:nphoton} 
\end{split}
\end{equation}
\refet{Here $\chi^{A(B)}_{ge/gf/ef}$ are the cavity shifts induced by qutrit A(B), and} $\epsilon_C$ is the amplitude of the cavity probe with $\epsilon_C=\kappa \sqrt{n}/2$. 
The qutrit operators are defined similarly to the qubit case, where $\sigma_+^{ge}=|e\rangle\langle g|$, $\sigma_-^{ge}=|g\rangle\langle e |$, $\sigma_+^{gf}=|f\rangle\langle g|$, $\sigma_-^{gf}=|g\rangle\langle f |$, $\sigma_+^{ef}=|f\rangle\langle e|$, and $\sigma_-^{ef}=|e\rangle\langle f |$. Thus we have $\sigma_x^{ge/ef/gf}=\sigma_+^{ge/ef/gf} + \sigma_-^{ge/ef/gf}$, $\sigma_y^{ge/ef/gf}=i\left(\sigma_+^{ge/ef/gf} - \sigma_-^{ge/ef/gf}\right)$,
as well as $\sigma_z^{gf}=-|g\rangle\langle g|+ |f\rangle\langle f|$. 
\refet{Here we choose $n=3$. The parameters we used for the cavity-qutrit interaction term and the cavity linewidth are shown in the first line of Table~\ref{tabs1}.} 
The $T_1$s and $T_2$s are set to optimistically large values of 500 $\mu$s, so that these decay channels contribute negligibly to the dynamics. The Rabi frequencies for the ``0 photon drive'' and the ``$n$ photon drive'' are chosen as $\Omega^{(0)} = \kappa /2$ and $\Omega^{(n)} = \kappa $ for optimization.

With the protocol applied to a 1-D qutrit chain containing $N$ qutrits, the Hamiltonian terms become
\begin{equation}
\begin{split}
\mathrm{\hat{H}}&_\mathrm{system}  = \hbar \sum_{i=1} ^ {N-1} \left( \frac{\chi_{gf}^i}{2}\sigma_z^{i,gf} + \frac{\chi_{ge}^i-\chi_{ef}^i}{2}|e\rangle^{i}\langle e |^{i} \right. \\
& \left. +\frac{\chi_{gf}^{\prime~i}}{2}\sigma_z^{i+1 ,gf} + \frac{\chi_{ge}^{\prime~i}-\chi_{ef}^{\prime~i}}{2}|e\rangle^{i+1}\langle e |^{i+1} \right) \hat{a}_i ^ \dagger \hat{a}_i,
\\
\end{split}
\end{equation}
\begin{equation}
\begin{split}
\mathrm{\hat{H}}_\mathrm{probe}=2\hbar \sum_{i=1} ^ { N-1} \epsilon_C^i \cos{\left(\frac{\chi^i_{gf}+\chi^{\prime~i}_{gf}}{2} t \right)  }\left(\hat{a}_i+\hat{a}^\dagger_i\right),
\end{split}
\end{equation}
\begin{equation}
\begin{split}
\mathrm{\hat{H}}_0=\hbar\Omega^{(0)}  \sum_{i=1} ^ { N} \left(\sigma_x^{i,ge}+\sigma_x^{i,ef}\right),
\end{split}
\end{equation}
and,
\begin{equation}
\begin{split}
&\mathrm{\hat{H}}_n  =  \hbar \sum_{i=1} ^ { N-1} 
 \Omega^{(n)}_i (-1)^{i-1} \left( \cos\left(n\frac{\chi^i_{gf}
 +\chi^{\prime~i}_{gf}}{2}t\right)
 (\sigma_x^{i,gf}-\right.
 \\
 &\sigma_x^{i+1,gf}) 
\left.-\sin \left(n\frac{\chi^i_{gf}+\chi^{\prime~i} _{gf}}{2}t\right)(\sigma_y^{i,gf}-\sigma_y^{i+1,gf}) \right).
\end{split}
\end{equation}
Here $\chi^i_{ge/gf/ef}$ and $\chi^{\prime~i}_{ge/gf/ef}$ are the cavity shifts on the $i_{th}$ cavity induced by the $i_{th}$ and the $(i+1)_{th}$ qutrit, and $\hat{a}_i(\hat{a}^\dagger_i)$ is the annihilation(creation) operator for the $i_{th}$ cavity. With $\kappa^i$ being the cavity linewidth of the $i_{th}$ cavity, we apply the probe strength for this cavity $\epsilon_C^i=\kappa^i \sqrt{n}/2$.
The terms $ \sigma_{x,y,z} ^ {i,ge/ef/gf} $ are the qutrit matrices for the $i_{th}$ qutrit. The ``0 photon drive'', $\Omega^{(0)}$, is chosen to be \refet{related to the averaged cavity linewidth as} $\sum_{i=1}^{N-1} \kappa ^i/2(N-1)$ and the ``$n$ photon drive'' \refet{applied on the $i_{th}$ cavity,} $\Omega^{(n)}_i$, is chosen to be $\kappa^i$. \refet{The averaged photon number $n$, is chosen to be $n=3$ up to optimization.}
The qutrit-cavity parameters and cavity linewidths are assumed to be the same for each cavity, equivalent to the two-qutrit case, which is shown in the first line of Table~\ref{tabs1}. 
\refet{The $T_1$ and $T_2$ settings are also the same as in the two-qutrit case unless otherwise specified.}

For the qutrit number $N_\mathrm{sites}\leq3$, we use the Qutip master equation solver to obtain the time evolution of the expectation value for the AKLT subspace projector. For $N_\mathrm{sites}=4$, the Monte Carlo solver is chosen for its better performance the in case of large dimensional Hilbert spaces. In the latter solver, the equivalence to the system evolution under the master equation is obtained by stochastically calculating the trajectories for quantum jumps.
The parameter settings are shown in the first line of Table~\ref{tabs1}, with a general ratio between $\chi_{ge}$ and $\kappa$ around 20.

\begin{table*}[h!]
\centering
\begin{tabular}{||l |c c c c c||}

 \hline
 &$\chi_{ge}/2\pi \ \mathrm{(MHz)}$ & $\chi_{gf}/2\pi\ \mathrm{(MHz)}$ & $\chi_{ge}^\prime/2\pi\ \mathrm{(MHz)}$ & $\chi_{gf}^\prime/2\pi\ \mathrm{(MHz)}$ & $\kappa/2\pi\ \mathrm{(MHz)}$ \\ [0.5ex] 
 \hline\hline
 Base parameters used in all simulations&40.00 & 79.20 & 38.00 & 76.00 & 2.00\\ 
 \hline\hline
 Ideal ``target'' parameters&40.00 & 80.00 & 40.00 & 80.00 & 2.00 \\
 \hline\hline
 Cavity 1 mismatched 10\% from target&43.64 & 87.81 & 39.67 & 87.81 & 1.93 \\
 \hline
 Cavity 2 mismatched 10\% from target&36.66 & 79.83 & 39.87 & 79.83 & 1.99 \\
 \hline
 Cavity 3 mismatched 10\% from target&41.52 & 76.96 & 42.79 & 76.96 & 1.94 \\
 \hline
 
\end{tabular}
\caption{\label{tabs1} {\bf The cavity-qutrit interaction parameters and the cavity linewidth for one cavity in the chain}. Line 1: The base parameters we choose for simulating two-qutrit as well as multi-qutrit protocol performance in the article, except in Fig.~\ref{fig5}(b). This set of base cavity parameters already \refeo{exhibit a small (5$\%$) mismatch} from the ideal ``target'' parameters.  Line 2: The ideal device parameters as the target of device fabrication processes. This set of parameters is the reference for generating the mismatching parameter. Line 3-5: The mismatched cavity parameters for a four-qutrit chain with its three cavities, the parameter deviations are randomly generated within 10 percent deviation from the ideal values. Such deviations are enlarged proportionally for generating the 20$\%$, 30$\%$, and 40$\%$ mismatching parameters as given in Fig.~\ref{fig5}(b). }
\end{table*}

Table \ref{tabs1} displays the dispersive shifts and cavity linewidths used for the simulations. To ensure that our results do not hinge on perfect parameter matches, we performed all simulations with ``base parameters'' that were near to what might be considered ideal. The base parameters and the ``target'' parameters are given in the first two lines of the table.  The base parameters are used for the simulations displayed Fig.~\ref{fig3}, Fig.~\ref{fig4}, and Fig.~\ref{fig5}(a), (c), (d). In Fig.~\ref{fig5}(b) we display simulation results where the neighboring qutrits and cavities have mismatched parameters. With an overall control of the Josephson inductance, we can assume that  perfect matching between $\chi_{gf}^i$ and $\chi_{gf}^{\prime~i}$ can be achieved for each cavity $i$ and its two coupled qutrits $i$ and $i+1$, via flux tuning on the qutrits. Other parameters, including the qutrit-cavity interaction term and the cavity linewidth, are mismatched. These parameters are given in lines 3-5 of Table \ref{tabs1}. For larger mismatches displayed in Fig.~\ref{fig5}(b), the deviations are simply scaled accordingly.


For optimization of the simulation process, it is desirable to make a cutoff at the maximal cavity photon population at the lowest value possible while maintaining accurate results.
As is presented in Fig.~\ref{figs2}, we thus monitored the cavity photon number population throughout the same stabilization process shown in Fig.~\ref{fig3}. 
With the ground state or the maximally mixed state unidirectionally projected into the $S_\mathrm{total} \in \{0,1\}$ subspace, the cavity photon number ramps up in about the first \refeo{500~ns} and then decays monotonically. 
Whichever initial state we chose, the cavity photon numbers for $n\geq4$ are quite small throughout the stabilization process. 
Actually, the \refeo{behavior} of cavity photon number $n\in \{0,1,2,3\}$ makes up 90 percent of the cavity state population, thus enabling a representative description of the overall system behavior with limited photon numbers. 
Consequently, we make a reasonable cutoff of the cavity photon population $n \leq 3$, with which the simulation results are shown in the insets of Fig.~\ref{figs2}(a), (b). 
By comparing the insets of Fig.~\ref{figs2}(a), (b) (with $n\leq 3$)  to the main panels (with $n\leq 9$) we see very similar photon number dynamics further confirming that a simulation cutoff of $n\leq 3$ produces accurate results.

\renewcommand{\thefigure}{S3}
\begin{figure}
\centering
\includegraphics[width = 0.5\textwidth]{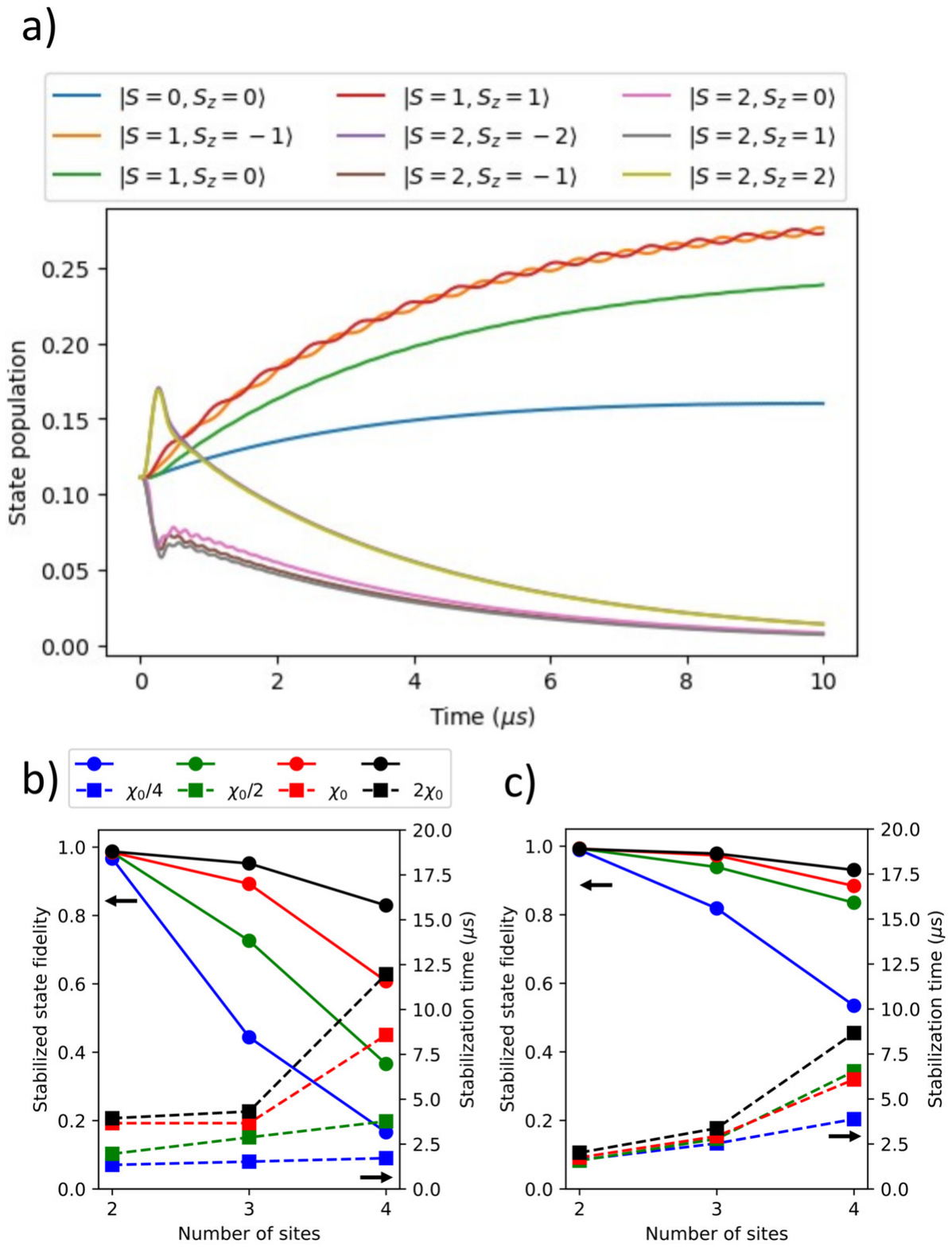}

\caption{\label{figs3}  {\bf Performance comparison for the protocol with different ``$n$ photon drives''}. (a) For the stabilization of one adjacent pair of qutrits initially prepared with fully mixed state with an alternative drive given in Eq.~\ref{eq:altdrive}. (b,c) The performance of the protocol with system size scaling up in terms of stabilization time and fidelity.  In panel (b) the rotation is $\hat{R}_\mathrm{eff}^n=\hat{S}_x^\mathrm{A} \otimes I^\mathrm{B}-I^\mathrm{A} \otimes \hat{S}_x^\mathrm{B}$ while in panel (c) the rotation is
$\hat{R}_\mathrm{eff}^n=\hat{R}_{gf}^\mathrm{A} \otimes I^\mathrm{B}-I^\mathrm{A} \otimes \hat{R}_{gf}^\mathrm{B}$. Here the cavity shifts scale as $\chi=\chi_0/4$ (blue), $\chi=\chi_0/2$ (green), $\chi=\chi_0$ (red) and $\chi=2\chi_0$ (black).
}
\end{figure}

\section{Alternative ``$n$ photon drives''} \label{app:d}

The ``$n$ photon drive'' serves to drive states back into the ALKT subspace. \refeo{The drives are activated when there are $n$ photons in the cavity. This drive can be chosen as any operator that has components rotating between the two subspaces inside and outside $S_\mathrm{total}=2$.} 
With different choices of the $n$ photon drive, the stabilization can in theory be accomplished with a varied converging time. For example, one alternative to the second line in Eq.~\ref{eqn:rot} could be
\begin{equation}
\hat{H}_\mathrm{eff}^n=\hat{S}_x^\mathrm{A} \otimes I^\mathrm{B} - I^\mathrm{A} \otimes \hat{S}_x^\mathrm{B}. \label{eq:altdrive}
\end{equation}
This alternative protocol is shown in Fig.~\ref{figs3} for the two-qutrit case as well as its scaling performance. Applied to two qutrits (Fig.~\ref{figs3}(a)), the protocol still effectively stabilizes the target subspace, but with a different distribution of states in the four-fold target subspace. However, by comparing Fig.~\ref{figs3}(b) (with the rotation Eq.~\ref{eq:altdrive}) and Fig.~\ref{figs3}(c) (with the rotation Eq.~\ref{eqn:rot}) we see that the choice of Eq.~\ref{eqn:rot} performs better at a larger number of sites.

\section{Potential experimental layout} \label{app:e}

\renewcommand{\thefigure}{S4}
\begin{figure}
\centering
\includegraphics[width = 0.5\textwidth]{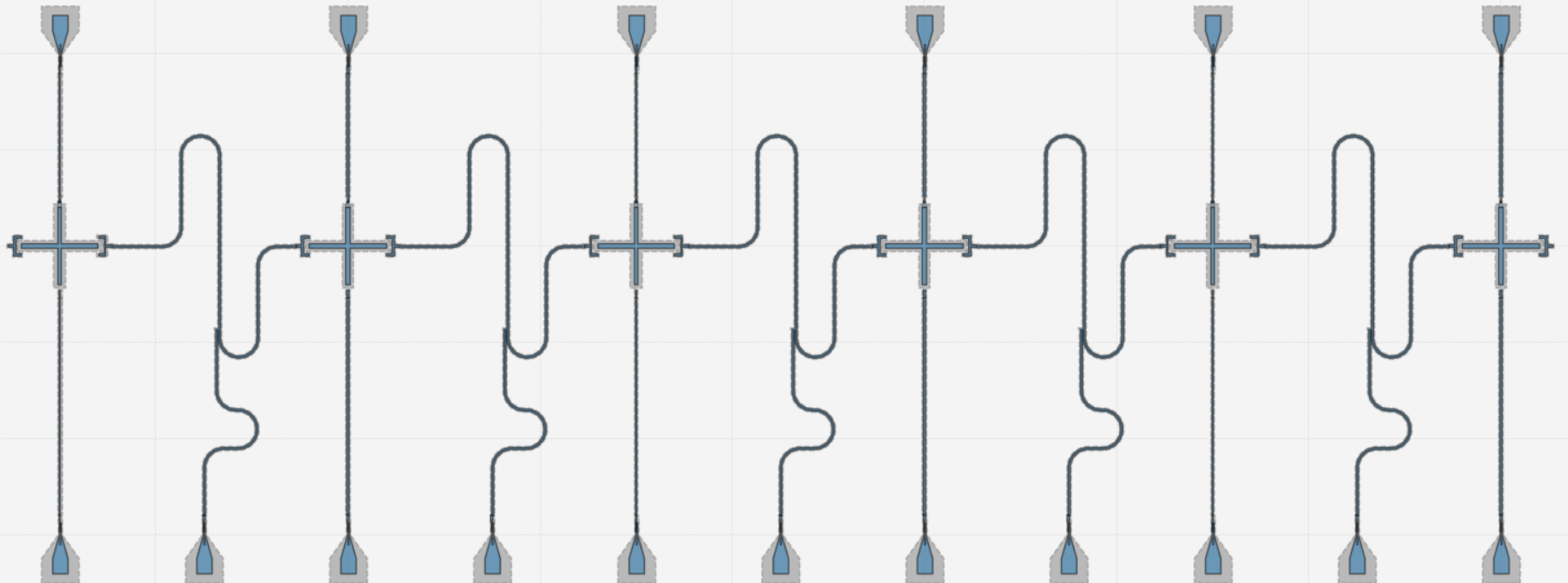}

\caption{\label{figs4}  {\bf Potential experimental device}. Sample layout of a $N=6$ qutrit chain, where dissipative coupling between neighboring transmon circuits is mediated by resonators.  Individual lines allow for flux tuning and microwave pulse control of each transmon.   }
\end{figure}

We propose an experimental design that can be realized with state-of-the-art fabrication capabilities, as a proof of principle of our scheme. This is shown in Fig.~\ref{figs4}. Each qutrit is attached to a flux line and a control line, where the qutrit frequencies can be tuned and the qutrit rotations can be applied. For the microwave cavity coupled to each adjacent pair of qutrits, there is a drive line to apply probes to the shared cavity. The shared cavities can also be utilized for state readout to perform quantum state tomography.

\renewcommand{\thefigure}{S5}
\begin{figure}
\centering
\includegraphics[width = 0.5\textwidth]{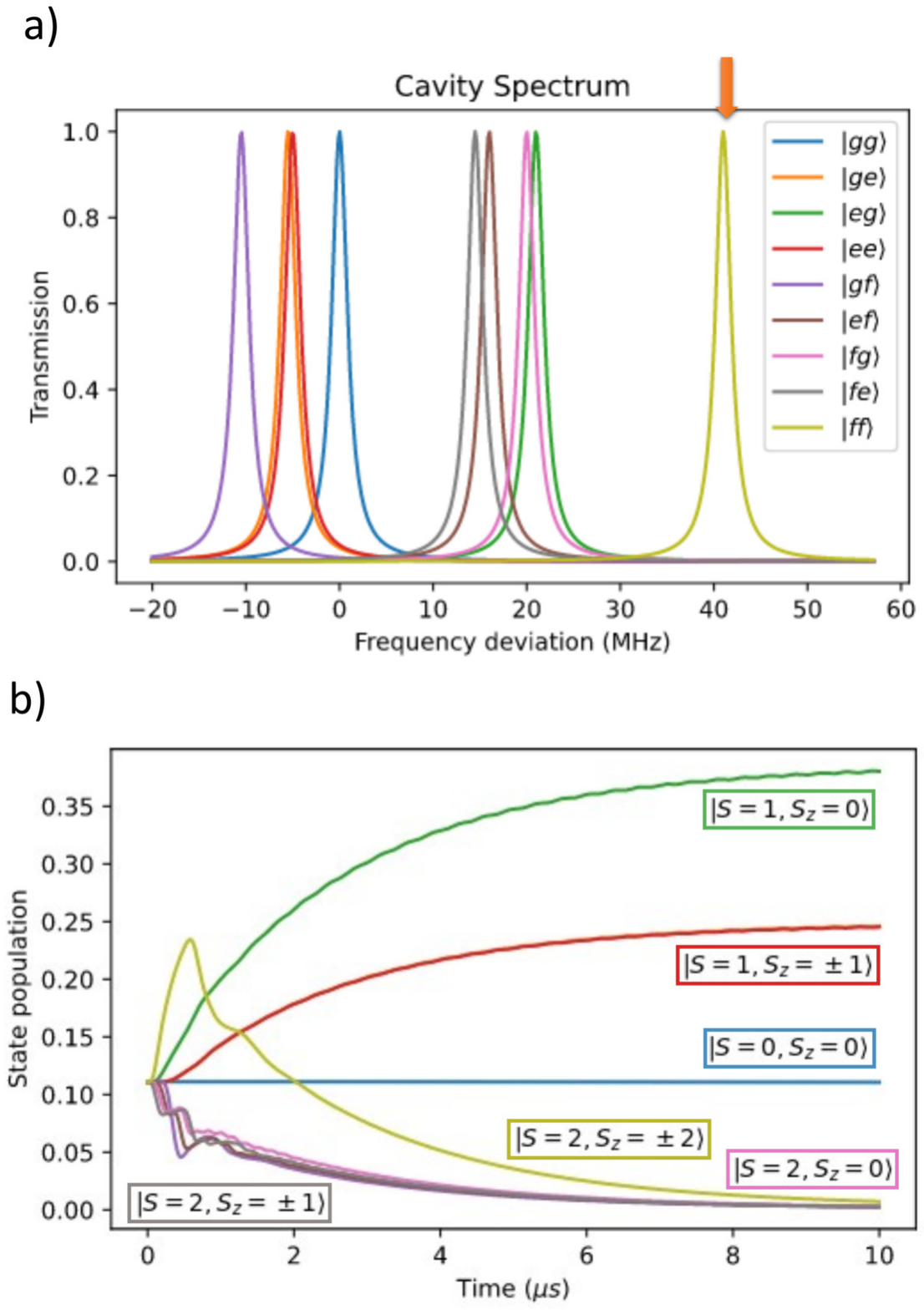}
\caption{\label{figs5}  {\bf Cavity probe and corresponding two-qutrit stabilization in the straddling regime.} \refeo{(a) A diagram of the cavity spectrum that can be obtained in the straddling regime, where cavity shift of state $|f\rangle$ is positive and the value is negative for state $|e\rangle$. In this case, the cavity probe at the $|gg\rangle$ peak would induce a large amount of extra dephasing, so only the $|ff\rangle$ peak is driven. This spectrum is diagrammatic with certain shifts and widths exaggerated to show the structure. (b) The two-qutrit protocol performance under the single-sided cavity probe. The protocol still efficiently drives the two-qutrit state into the 4-dimensional target subspace. Here since the cavity drive is single-sided, we adjust the ``0-photon drive'' frequency of the qutrits to be shifted with a small photon number induced by the single-side cavity probe.  }}
\end{figure}

\refet{In the main text, we analyzed an idealized set of parameters that yields the clearest setup for pedagogical reasons. Here we provide physical parameters to realize the high value of $\chi_{gf}$ required for high-fidelity operation of the protocol. We first introduce the straddling regime \cite{Koch2007} for the dispersive shift between superconducting transmon and the cavity. As is shown in Appendix~\ref{app:a}, $\chi_{gf}=\chi_2-\chi_0=\chi_{1,2}-\chi_{2,3}+\chi_{0,1}$. Since $\chi_{i,i+1}=g_{i,i+1}^2 /(\omega_{i,i+1}-\omega_r)$, and the coupling strengths have $g_{i,i+1}=\sqrt{i+1}g_0$, we have} 
\begin{equation}
\chi_{gf}=g_0^2(\frac{2}{\omega_{1,2}-\omega_r}-\frac{3}{\omega_{2,3}-\omega_r}+\frac{1}{\omega_{0,1}-\omega_r}).
\end{equation}
Since for the transmon energy level, we have \cite{Koch2007},
\begin{equation}
E_m \simeq -E_J+\sqrt{8E_CE_J}(m+\frac{1}{2})-\frac{E_C}{12}(6m^2+6m+3)
\end{equation}
\refet{Then $E_{01}=\sqrt{8E_CE_J}-E_C$, $E_{12}=\sqrt{8E_CE_J}-2E_C$, $E_{23}=\sqrt{8E_CE_J}-3E_C$. For $E_C/h = 400$ MHz $= \alpha/2\pi$, and $\sqrt{8E_CE_J}/h= 7$ GHz, we have reasonable qubit frequency and anharmonicity, and a good ratio of $E_J/E_C\sim 40$ that enables less fluctuations on the $|f\rangle$ state. When we have $\omega_{1,2}-\omega_r = \alpha/2$, then $\omega_{0,1}-\omega_r = 3\alpha/2$ and $\omega_{2,3}-\omega_r = -\alpha/2$, and thus $\chi_{ef}=32g_0^2/3\alpha $. To obtain $\chi_{ef}/2\pi=80$ MHz, we can have $g_0/2\pi = 55$ MHz which still stays in the dispersive limit. In this case, though, the cavity spectrum peaks are arranged in a different way since}
\begin{equation}
\chi_{ge}=\chi_1-\chi_0= g_0^2(\frac{2}{\omega_{0,1}-\omega_r}-\frac{2}{\omega_{1,2}-\omega_r}),
\end{equation}
\refet{where $\chi_{ge}=-8g_0^2/3\alpha$ is a negative value, $\chi_{ge}/2\pi=-20$ MHz. 
Thus, a slightly new driving strategy should be adopted. A diagram of the cavity spectrum is shown in Fig.~\ref{figs5} (a), where we probe the cavity only at the $|ff\rangle$ peak. The corresponding protocol performance is shown in Fig.~\ref{figs5} (b), and the multi-qutrit protocol performance is shown in Fig.~\ref{figs6}. Here, we show that the good isolation of the $|ff\rangle$ peak is achievable in realistic physical devices, and the consequential high performance can be expected. }

\renewcommand{\thefigure}{S6}
\begin{figure}
\centering
\includegraphics[width = 0.5\textwidth]{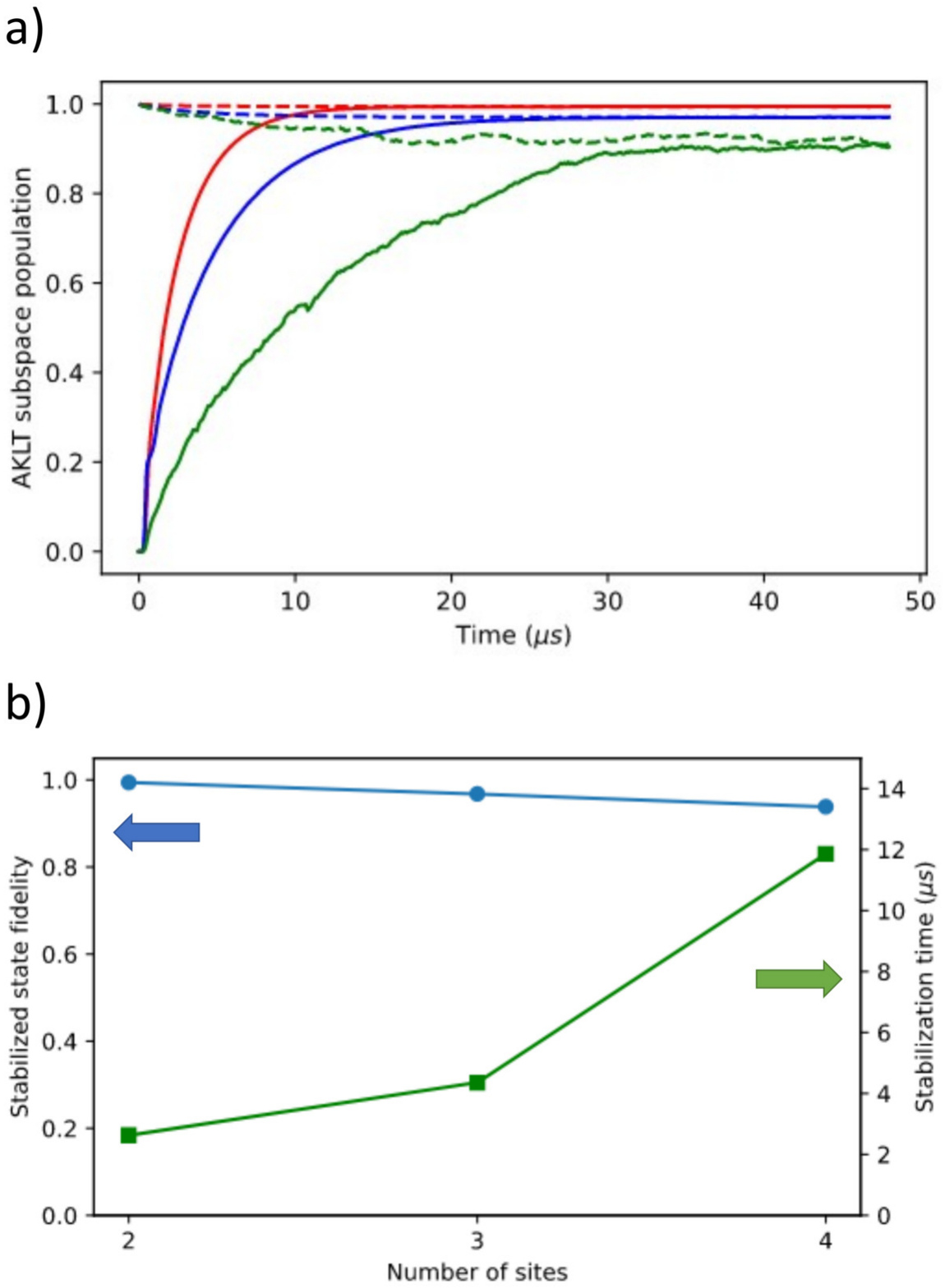}
\caption{\label{figs6}  {\bf Protocol performance with multiple qubits in the straddling regime.} \refeo{(a) The AKLT subspace population starting from the qutrits' ground state $\ket{g\ldots g}$ (solid colored lines) and starting within the AKLT subspace (dashed colored lines), we simulate the time evolution of the AKLT chain with open boundary conditions subspace population under the driven dissipative protocol with a varying number of qutrits, $N_\mathrm{sites} = 2$ (red), $N_\mathrm{sites} = 3$ (blue) and $N_\mathrm{sites} = 4$ (green) for comparison.
(b) Extracted fitting parameters as the final fidelity and the convergence time \refeo{with the same method as in Fig.~\ref{fig3} (b) on the total population in the AKLT subspace}. The blue dots represent the varied final fidelity with systems of two, three, and four qutrits (left axis), and the green dots represent the convergence time for the protocol (right axis). } }
\end{figure}


\end{document}